\documentclass[]{aa}
\usepackage{amsmath}
\usepackage{graphicx}
\usepackage{xcolor}
\usepackage{txfonts}
\usepackage{multirow}
\usepackage{natbib}
\titlerunning{5-10 keV XLF}
\authorrunning{S. Fotopoulou et al.}

\bibpunct{(}{)}{;}{a}{}{,} 

\begin{document}

    \title{The 5 -- 10 \lowercase{$\rm{ke}$}V AGN luminosity function at 0.01<z<4.0}

    \author{S. Fotopoulou\inst{1,2,3,4}
            \and J. Buchner\inst{2}
            \and I. Georgantopoulos\inst{5}
            \and G. Hasinger\inst{1,6}
            \and M. Salvato\inst{1,2}
            \and A. Georgakakis\inst{2,5}
            \and N. Cappelluti\inst{7}
            \and P. Ranalli\inst{5,7}
            \and L.T. Hsu\inst{2}
            \and M. Brusa\inst{2,7,8}
            \and A. Comastri\inst{7}
            \and T. Miyaji\inst{9,10}
            \and K. Nandra\inst{2}
            \and J. Aird\inst{11}
            \and S. Paltani\inst{4}}

    \institute{Max Planck Institut f\"ur Plasma Physik, Boltzmannstrasse 2, 85748 Garching Germany  
    \and Max Planck Institut f\"ur Extraterrestrische Physik, Giessenbachstrasse, 85748 Garching Germany 
    \and Technische Universit\"at M\"unchen, James-Franck-Strasse 1, 85748 Garching Germany 
    \and Department of Astronomy, University of Geneva, chemin d'Ecogia 16, CH-1290 Versoix, Switzerland \\ \email{Sotiria.Fotopoulou@unige.ch} 
    \and IAASARS, National Observatory of Athens, GR-15236 Penteli, Greece 
    \and Institute for Astronomy, University of Hawaii, 2680 Woodlawn Drive Honolulu, HI 96822-1839, USA 
    \and INAF-Osservatorio Astronomico di Bologna, via Ranzani 1, 40127, Bologna, Italy 
    \and Dipartimento di Fisica e Astronomia, Universit\`a di Bologna, viale Berti Pichat 6/2, 40127 Bologna, Italy 
    \and Instituto de Astronomía, Universidad Nacional Autónoma de México, Ensenada, Baja California, Mexico 
    \and University of California San Diego, Center for Astrophysics and Space Sciences, 9500 Gilman Drive, La Jolla, CA 92093-0424, USA 
    \and Institute of Astronomy, University of Cambridge, Madingley Road, Cambridge CB3 0HA, UK 
    }

    \date{Received; accepted}

    \abstract
        {The active galactic nuclei X-ray luminosity function traces actively accreting supermassive black holes and is essential for the study of the properties of  the active galactic nuclei (AGN) population,   black hole evolution, and galaxy-black hole coevolution.
        Up to now, the AGN luminosity function has been estimated several times in soft ($\rm{0.5-2\,keV}$) and hard X-rays ($\rm{2-10\,keV}$). AGN selection in these energy ranges often suffers from identification and redshift incompleteness and, at the same time, photoelectric absorption can obscure a significant amount of the X-ray radiation. We estimate the evolution of the luminosity function in the $\rm{5-10\,keV}$ band, where we effectively avoid the absorbed part of the spectrum, rendering absorption corrections unnecessary up to $\rm{N_H{\sim}10^{23}cm^{-2}}$.
        Our dataset is a compilation of six wide, and deep fields: MAXI, HBSS, XMM-COSMOS, Lockman Hole, XMM-CDFS, AEGIS-XD, Chandra-COSMOS, and Chandra-CDFS. This extensive sample of ${\sim} 1110$ AGN ($\rm{0.01<z<4.0}$, $\rm{41<\log L_x<46}$) is 98\% redshift complete with 68\% spectroscopic redshifts. For sources lacking a spectroscopic redshift estimation we use the probability distribution function of photometric redshift estimation specifically tuned for AGN, and a flat probability distribution function for sources with no redshift information. We use Bayesian analysis to select the best parametric model from  simple pure luminosity and pure density evolution to more complicated luminosity and density evolution and luminosity-dependent density evolution. We estimate the model parameters that describe best our dataset  separately for each survey and for the combined sample.
        We show that, according to Bayesian model selection, the preferred model for our dataset is the luminosity-dependent density evolution (LDDE). Our estimation of the AGN luminosity function does not require any assumption on the AGN absorption and is in good agreement with previous works in the $\rm{2-10\,keV}$ energy band based on X-ray hardness ratios to model the absorption in AGN up to redshift three. Our sample does not show evidence of a rapid decline of the AGN luminosity function up to redshift four.}

    \maketitle


\section{Introduction}
\label{sec:introduction}
    The X-ray luminosity function (XLF) of active galactic nuclei (AGN) and its evolution provides a view of black hole (BH) growth across cosmic time. Several studies use the XLF to constrain models of BH evolution through simulations and semi-analytic models \citep[e.g.,][]{Mahmood2005, Hopkins2005a,Hirschmann2012,Hirschmann2014,Enoki2014} and to investigate the possible galaxy-BH coevolution \citep[e.g.,][]{Hopkins2007,  Marulli2008, Zheng2009, Fanidakis2011}. The XLF  is also used to constrain the properties of the AGN population, for example, by creating population synthesis models that describe the cosmic X-ray background (CXB) and thus inferring the fraction of Compton-thick AGN \citep[e.g.,][]{Gilli2007, Treister2009, Draper2009, Akylas2012}. Additionally, the XLF is used to test the still open question of AGN triggering: mergers vs secular processes \citep{Draper2012}.  The multivariate changes of the AGN phase can be studied by combining the XLF with luminosity functions in other wavelengths  \citep[e.g.,][]{Han2012, Hopkins2005b}. For example, the connection between X-ray and infrared radiation from AGN has been studied by means of reconciling the CXB with the infrared background through the corresponding luminosity functions \citep{Ballantyne2007}.

    Early X-ray surveys showed that the space density of AGN follows a broken power-law distribution and it was proposed that the XLF only evolves with redshift in luminosity \citep[pure luminosity evolution, hereafter PLE;][]{Maccacaro1983, Maccacaro1984}, while subsequent studies showed that the evolution stops, or dramatically slows down after a critical redshift value \citep[e.g.][]{Boyle1994,Page1996,Jones1997}. Recent works in the soft X-ray regime ($\rm{0.5-2\,keV}$) support a luminosity-dependent density evolution (LDDE) over the simple PLE with the number density of AGN peaking in redshift $\rm{z=1-2}$ \citep{Miyaji2000, Hasinger2005, Ebrero2009}. Similarly in the hard X-ray band ($\rm{2-10\,keV}$) several works support LDDE over PLE \citep{Ueda2003, LaFranca2005, Ebrero2009, Ueda2014, Miyaji2015}. Other studies at the same energy band also tested simultaneous variations in luminosity and density with different parametrization, namely, independent luminosity and density evolution \citep[ILDE;][]{Yencho2009} and luminosity and density evolution  \citep[LADE;][]{Aird2010}.

    According to the unified model of AGN \citep{Antonucci1993, Urry1995}, a supermassive black hole is found at the center of each AGN, surrounded by an accretion disk and a torus of gas and dust. The current accepted view of the radiation mechanism includes X-ray production in the vicinity of the black hole through Comptonization of disk photons in a population of hot thermal electrons \citep{Haardt1993}. The torus is responsible for the obscuration of  X-rays due to photoelectric absorption. The torus also generates infrared emission, which is  optical radiation reprocessed by the dust. For the luminosity function in the X-ray energy ranges $\rm{0.5-2\,keV}$ and $\rm{2-10\,keV,}$ a correction factor must be applied depending on the absorption power of the obscuring torus in each source. This absorption power is either calculated from the spectrum or roughly estimated from the observed flux in at least two X-ray energy bands. An additional correction factor is often applied to take redshift incompleteness into account in the sample under investigation. 

    In this work, we study X-ray emission in a hard band ($\rm{5-10\,keV}$), avoiding the absorbed part of the spectrum. Previous works over the same energy range attempted to put constraints on the bright end of the $\rm{5-10\,keV}$ XLF. \citet{LaFranca2002} used ${\sim}160$ AGN detected in BeppoSAX, ASCA, and HEAO 1 with a flux limit of $\rm{F_x=3\times10^{-14}erg\,s^{-1}}\,cm^{-2}$, while \citet{Ebrero2009} used a sample of ${\sim}120$ XMM detected AGN with a flux limit of $\rm{F_x=6.8\times10^{-15}erg\,s^{-1}}\,cm^{-2}$. The small number of available AGN was not enough to differentiate between PLE and LDDE models and the authors resorted to fixing the evolutionary parameters of the luminosity function. Similarly, efforts have been made to  determine the X-ray luminosity function at energies higher than {$\rm{10\,keV}$} using objects detected with {\it INTEGRAL} and {\it SWIFT}-BAT \citep{Beckmann2006,Sazonov2007,Paltani2008,Burlon2011}. These samples include very bright local objects (z<0.1), which put constraints on the local luminosity function, but are not able to constrain the evolution of the AGN number density with redshift.

With the combination of recent multiwavelength surveys, we are able to create a sizable sample of ${\sim}1100$ sources with a flux limit of about $\rm{F_x=1.5\times10^{-16}erg\,s^{-1}\,cm^{-2}}$, which is 10 times deeper and 5-6 times more numerous than previously available samples in the same energy range. Spectroscopic redshifts (spec-z), combined with accurate photometric redshifts (photo-z), provide a 98\% redshift complete sample that is ideal for probing AGN evolution. We determine the XLF and its evolution in the $\rm{5-10\,keV}$ band testing the evolutionary models used in the literature. We perform Bayesian analysis to investigate in detail the probability distribution function for each parameter that describes the luminosity function, per survey and in the combined sample. This approach provides an accurate view of the parameters without assuming Gaussian distribution around the best value. In \S \ref{sec:data} we present the surveys used for this work, and in \S \ref{sec:510} we discuss the selection of the $\rm{5-10\,keV}$ energy range.  In \S \ref{sec:LF} we describe the models we used in estimating the luminosity function. In \S \ref{sec:results} we describe our analysis and the selection of the best model for our dataset. In \S \ref{sec:discuss} we discuss the LDDE model, compare the information contained in each survey separately, and predict the expected number of AGN for future surveys. We adopt the cosmological parameters ${{\Omega}_{\Lambda}=0.7}$, ${{\Omega}_m=0.3}$, and $\rm{H_0=70\,kms^{-1}Mpc^{-1}}$.


\section{Dataset definition}
\begin{figure*}
    \begin{tabular}{cc}
    \includegraphics[width=0.5\linewidth]{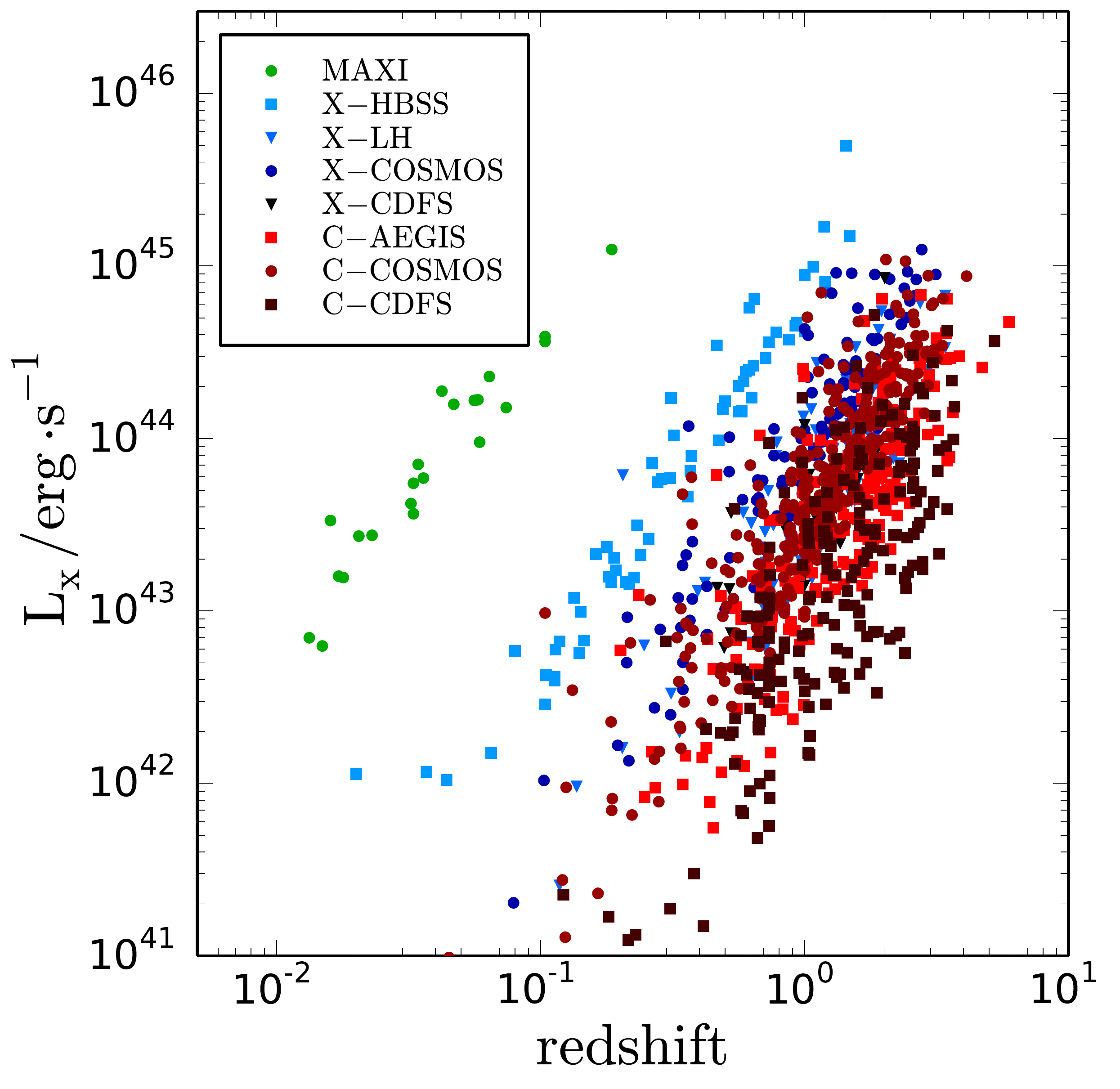} & \includegraphics[width=0.5\linewidth]{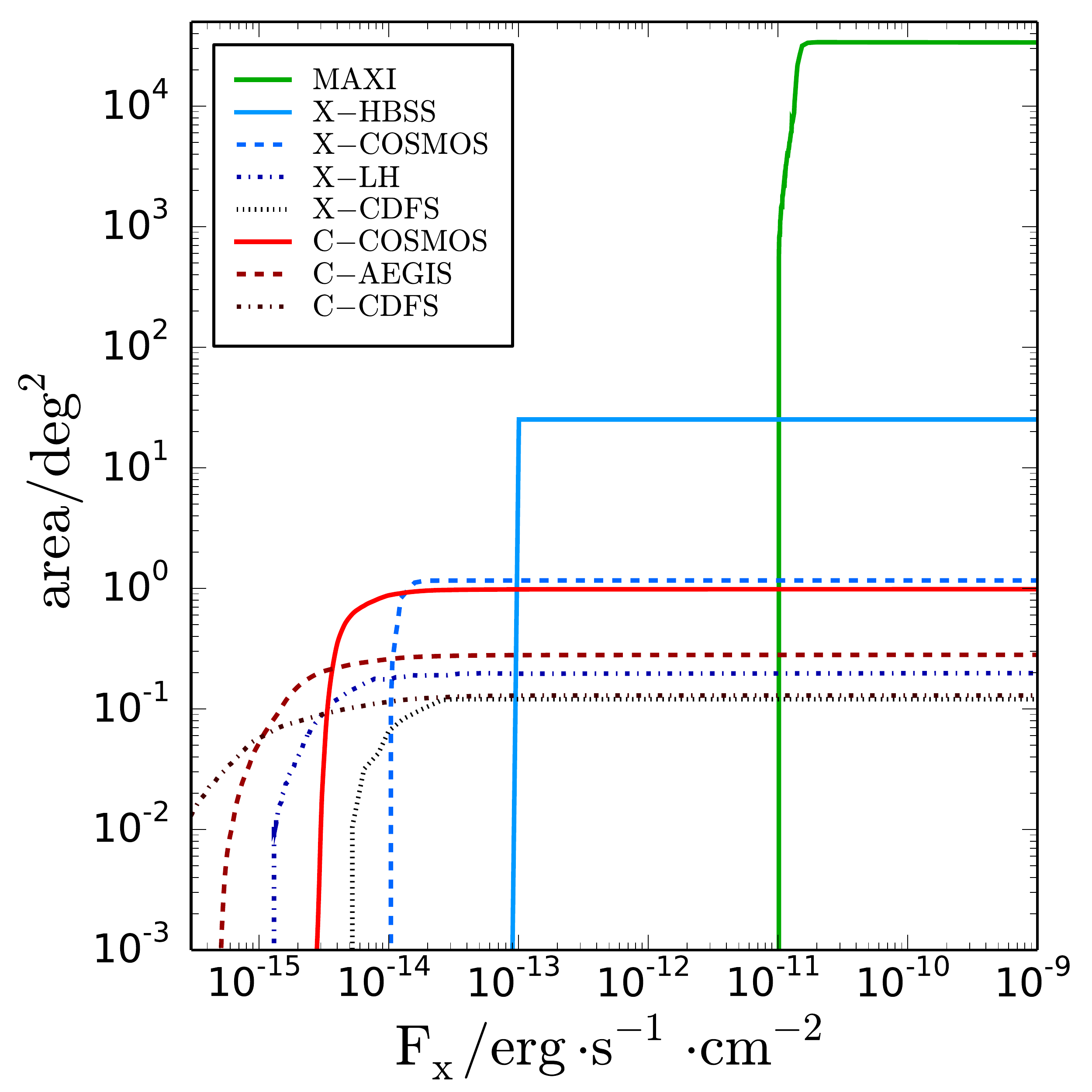} \\
    (a) & (b) \\
    \end{tabular}
    \caption{(a) $\rm{5-10\,keV}$ luminosity - redshift plane, showing sources with either spectroscopic or photometric redshift estimation. (b) Sky coverage as a function of X-ray flux for each survey.
    \label{fig:acurve}}
    \end{figure*}

   We combined wide/medium angle surveys (MAXI, HBSS, COSMOS) with pencil beam X-ray fields (Lockman Hole (LH), AEGIS, CDFS) to create a sample of $\rm{5-10\,keV}$ detected AGN. We exclude  known stars, galaxies, and galaxy clusters from these samples. The combined sample consists of 1115 AGN with luminosities between $\rm{10^{41}-10^{47}\,erg\,sec^{-1}}$ in the redshift range $\rm{0.01<z<5}$. As can be seen from Fig 1 (a), about ten sources have luminosity between 41<logL<42; this amounts to roughly 1\% of our sample. We have tested that using a limit of $\rm{10^42\,erg\,sec^{-1}}$ or $\rm{10^{41}\,erg\,sec^{-1}}$ does not significantly alter the estimation of the luminosity function parameters. Good coverage of the luminosity - redshift plane (Fig. \ref{fig:acurve}a) was possible as a result of the wide range of sky coverage and X-ray depth reached with the combination of these fields (Fig. \ref{fig:acurve}b). 

The sample is 98\% complete in redshift, and 68\% of the redshifts is spectroscopically determined. The remaining 29.7\% of the sample consists of sources from the fields XMM-COSMOS, Lockman Hole, AEGIS, and Chandra Deep Field South, where the multiwavelength coverage of the fields facilitated the computation of accurate photometric redshifts. The accuracy in all fields is better than $\rm{\sigma_{NMAD}}=0.07$ with a small fraction of outliers \citep{Salvato2009,Salvato2011,Fotopoulou2012,Hsu2014,Nandra2015,Aird2015}. All photometric redshifts were obtained with the SED fitting code LePhare\footnote{http://www.cfht.hawaii.edu/~arnouts/LEPHARE/lephare.html}, and we use the redshift probability distribution function for our analysis.  The probability distribution function is proportional to $e^{-\chi_{min}^2(z)/2}$, where $\chi_{min}^2$ refers to the galaxy or AGN template with the minimum $\chi^2$ at a given redshift. This function is normalized to 1 to obtain the probability distribution function. We assign a flat probability distribution function in the range z=0-7 for sources for which no photometric
redshift determination is possible.
    \subsection{Surveys}\label{sec:data}
    \paragraph{\bf MAXI extragalactic survey}
                              The Monitor of All-sky X-ray Image (MAXI) mission on board the International Space Station (ISS) observes the entire sky every 92 minutes with two instantaneous fields of view covering each $\rm{160^{\circ}\times3^{\circ}}$. MAXI consists of two cameras, the Gas Slit Camera (GSC; \citealt{Sugizaki2011,Mihara2011}) sensitive in the $\rm{2-30\,keV}$ energy band and the Solid-state Slit Camera (SSC; \citealt{Tsunemi2010,Tomida2011}) sensitive in the $\rm{0.5-12\,keV}$ energy band.
               \citet{Hiroi2011} presented the first MAXI/GSC seven-month data catalog of sources detected in the $\rm{4-10\,keV}$ band and at high Galactic latitudes ($|b|>10^{\circ}$, $\rm{34,000deg^2}$). \citet{Ueda2011} used 37 AGN from this catalog to compute the local AGN luminosity function. Here we are using the 22 AGN out of the 37 AGN presented in \citet{Ueda2011} that have spectroscopic redshift $\rm{z>0.01}$. The flux limit\footnote{As a flux limit, we quote the flux of the faintest source in the sample.} of the sample is $\rm{F_x=1.1\times10^{-11}erg\,s^{-1}\,cm^{-2}}$ covering a redshift range $0.01<z<0.19$ with median redshift $\rm{z=0.034}$ and median luminosity $\rm{\log L_x=43.85}$.
    \paragraph{\bf Hard Bright Serendipitous Survey (HBSS)}
               The XMM-Newton Bright Serendipitous Survey covers 25 $\rm{deg^2}$ \citep{DellaCeca2004} and provides two flux limited samples in the $\rm{0.5-4.5\,keV}$ and $\rm{4.5-7.5\,keV}$ band. We are using the hard sample $\rm{4.5-7.5\,keV}$ with a flux limit of $\rm{F_x=1\times10^{-13}erg\,s^{-1}\,cm^{-2}}$. The optical counterparts and spectroscopic redshifts are presented in \citet{Caccianiga2008}. The sources cover the redshift range $0.02<z<1.48$ with median redshift $\rm{z=0.312}$ and median luminosity $\rm{\log L_x=43.8}$. This survey covers a large area introducing sample rare bright objects at higher redshift compared to MAXI. 
    \paragraph{\bf XMM-COSMOS}
                The XMM-COSMOS field is one of the widest contiguous XMM fields covering 2 $\rm{deg^2}$.  The XMM-COSMOS field contains 245 sources detected in the $\rm{5-10\,keV}$ \citep{Cappelluti2009}, with a flux limit of $\rm{F_x=5\times10^{-15}erg\,s^{-1}\,cm^{-2}}$  and   a good balance between depth and sky
coverage. The unprecedented multiwavelength coverage of this field provides optical \citep{Capak2007}, ultraviolet \citep{Zamojski2007}, near-infrared \citep{McCracken2010}, mid-infrared \citep{Sanders2007,Ilbert2009,Frayer2009} counterparts, and spectroscopic redshifts \citep{Trump2009,Lilly2007,Lilly2009} \citep[see][for a summary]{Brusa2007, Brusa2010}. For 79.2\% of the sample spectroscopic redshifts are available, while the remaining 18.3\% of the sample has high quality photometric redshifts \citep{Salvato2009,Salvato2011} reaching a total of 97.5\% redshift completeness. The X-COSMOS sources span a redshift range $0.04<z<3.14$, with a median redshift of $\rm{z=1.1}$ and median luminosity $\rm{\log L_x=44.05}$.
    \begin{table*}
    \begin{center}
    \caption{Redshift information per X-ray field.}
    \label{tab:fields}
    \begin{tabular}{lcccccccc} \hline
    \multirow{2}{*}{field} &  area & \multirow{2}{*}{No. sources} & \multirow{2}{*}{spec-z} & \multirow{2}{*}{photo-z} & no & median & completeness & redshift \\           
     &  ($\rm{deg^2}$) & & & & redshift & redshift & completeness & range \\
    \hline

    MAXI                            & 33800     & 22      & 22      & 0     &  0   & 0.04       & 100\%           & 0.01-0.19 \\
    XMM-HBSS                        & 25        & 64      & 62      & 0     & 2    & 0.31       & 96.9\%         & 0.02-1.48 \\
    XMM-COSMOS\tablefootmark{a}     & 2.15/1.16  & 245/115 & 194/83  & 45/30 & 6/2  & 1.07/1.06  & 97.5\%/98.3\%  & 0.04-3.14 \\
    XMM-LH                          & 0.2       & 88      & 50      & 38    & 0    & 1.19       & 100\%          & 0.12-3.41 \\
    XMM-CDFS\tablefootmark{a}       & 0.26/0.12 & 137/30  & 109/18  & 24/8  & 4/4  & 1.31/1.22  & 97.1\%/86.7\%  & 0.12-3.8 \\
    Chandra-COSMOS                  & 0.98      & 357     & 257     & 87    & 13   & 1.21       & 96.4\%         & 0.04-4.1 \\
    Chandra-AEGIS-XD                & 0.28      & 244     & 133     & 111   & 0    & 1.51       & 100\%          & 0.04-5.9 \\
    Chandra-CDFS                    & 0.13      & 195     & 136     & 58    & 1    & 1.36       & 99.5\%         & 0.1-5.2  \\ \hline
    total sample\tablefootmark{b}   & 33828     & 1115    & 761     & 332   & 22   & 1.19       & 98\%          &  0.01-5.9   \\
    \hline
    \end{tabular}
    \end{center}
    \tablefoot{
    \tablefoottext{a}{Numbers refer to the full XMM field/not overlapping with Chandra area, respectively.}\\
    \tablefoottext{b}{Removing the overlapping area of XMM and Chandra observations in the COSMOS and CDFS fields as described in the text.}
    }
    \end{table*}
    \paragraph{\bf XMM - Lockman Hole (LH)}
                The Lockman Hole is one of the deepest XMM-Newton fields. The X-ray catalog presented in \citet{Brunner2008} contains 88 sources detected in the $\rm{5-10\,keV}$ band and the optical counterparts are presented in \citet{Rovilos2011} with flux limit $\rm{F_x=2\times10^{-15}erg\,s^{-1}\,cm^{-2}}$. Photometric redshifts are presented in \citet{Fotopoulou2012} reaching 98.8\% completeness. The Lockman Hole dataset provides faint sources detected with XMM, allowing us to probe the faint end of the luminosity function at redshift above 0.5. This dataset covers a redshift range of $0.118<z<3.41$ with median redshift $\rm{z=1.192}$ and median luminosity $\rm{\log L_x=43.78}$. 
    \paragraph{\bf XMM - Chandra Deep Field South (XMM - CDFS)}
        The Chandra Deep Field South has been observed by XMM for a total exposure of 3.5 Ms. We include the $\rm{5-10\,keV}$ detections presented in \citet{Ranalli2013}, where  the optical identification and spectroscopic redshift sample is also described. The XMM observations reach a flux limit of $\rm{F_x=8\times10^{-16}erg\,s^{-1}\,cm^{-2}}$. The photometric redshift described in \citet{Hsu2014} are specifically tuned for AGN and show improved accuracy over previous estimations for the same area. This dataset covers a redshift range of $0.5<z<2.5$ with median redshift $\rm{z=1.22}$ and median luminosity $\rm{\log L_x=43.58}$.     \paragraph{\bf Chandra-COSMOS}
    The COSMOS field has also been observed with the Chandra observatory, and  we use the $\rm{160\,ks}$ C-COSMOS observations of \citet{Elvis2009} covering an area of $\rm{{\sim}1.0\,deg^2}$. The reduction of the X-ray data is performed as described in \citet{Laird2009}, yielding 357 sources detected in the $\rm{4-7\,keV}$ energy band with flux limit $\rm{F_x=1.8\times10^{-15}erg\,s^{-1}\,cm^{-2}}$. The counterpart association and photo-z estimation are described in \citet{Aird2015}. The Chandra-COSMOS sources span a redshift range $0.04<z<4.1$ with a median redshift of $\rm{z=1.2}$ and median luminosity $\rm{\log L_x=43.75}$.
    \paragraph{\bf AEGIS-XD}
    The AEGIS-XD field spans an area of $\rm{0.2\,deg^2}$ and benefits from multiwavelength coverage. The X-ray detection was performed with Chandra in the $\rm{4-7\,keV}$ energy range, and the counts were transformed to flux in the $\rm{5-10\,keV}$ range, assuming that the X-ray spectrum is a power-law with $\Gamma=1.4$ \citep{Nandra2015}. The optical counterparts were retrieved with a likelihood ratio approach, using the Rainbow Cosmological Surveys Database \citep[][a,b]{Barro2011a}.
    Spectroscopic redshifts are a compilation from the DEEP2 \citep{Newman2012} and DEEP3 fields \citep{Cooper2012} and from the MMT \citep{Coil2009}. The AEGIS-XD survey is one of the two surveys with large number of objects in our sample with twice the area of LH and XMM-CDFS. The flux limit of AEGIS-XD is $\rm{F_x=5.9\times10^{-16}erg\,s^{-1}\,cm^{-2}}$ and the sources cover the redshift range $0.05<z<3.85$ with median redshift $\rm{z=1.53}$ and median luminosity $\rm{\log L_x=43.64}$.
    \paragraph{\bf Chandra - Chandra Deep Field South (Chandra - CDFS)}
     The Chandra Deep Field South is the deepest to date X-ray field, covering an area of $\rm{0.1\,deg^2}$. We use the 4Ms observations \citep{Xue2011}, reduced as described in \citet{Laird2009}, similar to the Chandra-COSMOS observations. The counterpart assignment is described in \citet{Aird2015}, while the photo-z estimation are the same as for XMM-CDFS presented in \citet{Hsu2014}. The flux limit of Chandra - CDFS is $\rm{F_x=1.4\times10^{-16}erg\,s^{-1}\,cm^{-2}}$. The Chandra-CDFS contains 195 sources spanning a redshift range $0.1<z<5.2$ with a median redshift of $\rm{z=1.3}$ and median luminosity $\rm{\log L_x=43.2}$.

\vspace{0.2cm}
  \citet{Avni1980} introduced the coherent volume addition to  analyze independent samples simultaneously. The main idea is that a source is considered detectable in any of the surveys as long as it is above the detection threshold for that particular survey. From the samples gathered for this work, sources in COSMOS and CDFS appear twice when considering both the XMM and Chandra observations since the fields partially overlap on the sky. \citet{Miyaji2015} solved this issue by merging the detection lists and creating an effective area curve that describes both XMM and Chandra COSMOS observations. Here instead we  adopt a different strategy. Since both Chandra-COSMOS and Chandra-CDFS cover a smaller area than the respective XMM observations, we keep the entire Chandra field and select the XMM observations that do not fall in the Chandra area. With this approach, we have two independent fields profiting both from the depth of the Chandra observations and the wider area covered by XMM. In Table \ref{tab:fields} we gather the area coverage, number of detected sources, and redshift information for all fields.

    \subsection{ 5-10 keV selection}
    \label{sec:510}

\begin{figure}
    \includegraphics[width=\linewidth]{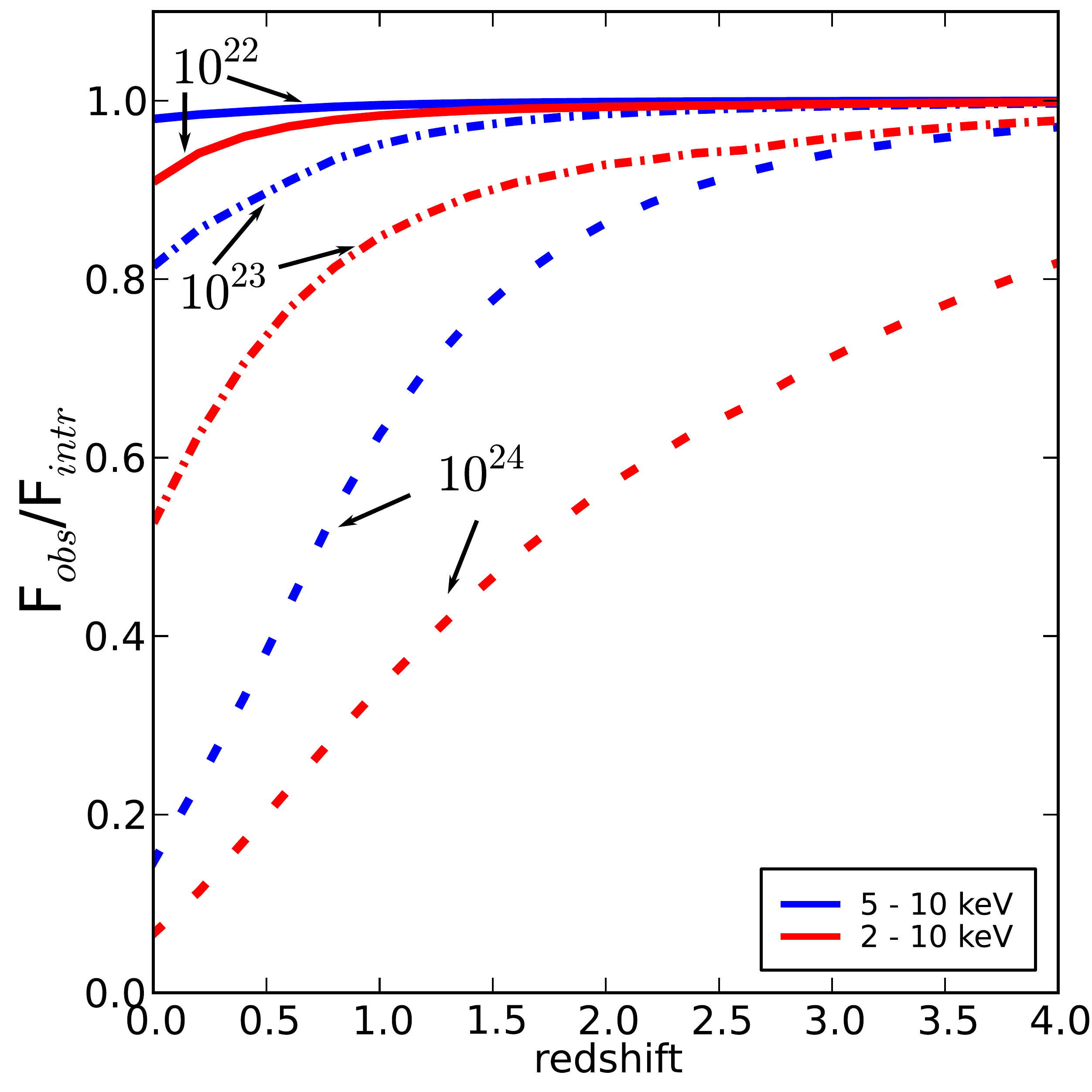}
    \caption{Observed flux over intrinsic flux for a power-law spectrum with photoelectric absorption in the $\rm{5-10\,keV}$ band (blue) and $\rm{2-10\,keV}$  energy band (red). The observed $\rm{5-10\,keV}$ flux is never lower than 80\% of the intrinsic at all redshifts column densities $\log N_H=23$.
    \label{fig:FluxRatio}}
    \end{figure}

    AGN X-ray spectra above $\rm{1\,keV}$ can be described adequately as a power-law distribution, $F(E)\propto E^{-\Gamma}$,
    with $\Gamma{\rm=1.4-1.6}$ for radio loud AGN and $\Gamma\rm{=1.8-2.0}$ for radio-quiet AGN \citep{Nandra1994,Reeves2000,Piconcelli2005,Page2005,Mateos2005}.
     We calculate theoretical flux and count rate values for different $\rm{N_H}$ and redshift values, assuming a power-law spectrum with photon index $\Gamma$ = 1.9 via models \emph{zphabs*cutoffpl} in PyXspec, the Python interface to
Xspec \citep{Arnaud1996}. In Fig. \ref{fig:FluxRatio} we plot the expected flux ratio for three $N_H$ values as a function of redshift. The blue lines correspond to the $\rm{5-10\,keV}$ energy band and the red lines correspond to the $\rm{2-10\,keV}$ energy band. We show that even for high absorbing column densities such as $\rm{10^{23}cm^{-2}}$, the observed flux in the $\rm{5-10\,keV}$ energy range is more than 90\% of the intrinsic for $\rm{z>1}$, and never less than 80\% even at lower redshifts.
In this plot we do not include the effect of Thompson scattering, which would reduce the flux up to 2.3\% and 18.8\% in the $\rm{2-10\,keV}$
band at $\rm{\log N_H=22}$ and $\rm{\log N_H=23,}$ respectively.

    XMM-Newton and Chandra are less sensitive to energies above $\rm{7\,keV}$. Therefore, only bright sources are detected in the $\rm{5-10\,keV}$ energy band since typically the detection limit in this band is an order of magnitude higher compared to the $\rm{0.5-2\,keV}$ energy band. Consequently, this work does not include very faint sources ($\rm{F_x<5.9\times10^{-16}erg\,s^{-1}\,cm^{-2}}$). Even though there are interesting cases on an individual source basis, we do not expect differences in a study of the global population as it has been demonstrated that the $\rm{5-10\,keV}$ does not select a special AGN population \citep{DellaCeca2008}.

    \section{Modeling the luminosity function}
    \label{sec:LF}
    Early observations of X-ray AGN showed that the local luminosity function ($\rm{z{\sim}0}$) is well described by a broken power-law distribution \citep{Maccacaro1983,Maccacaro1984}
    \begin{equation}\label{eq:LF0}
        \frac{d\phi(L,z=0)}{d\log{L}}=\frac{A}{\left(\frac{L}{L_0}\right)^{\gamma_1}+\left(\frac{L}{L_0}\right)^{\gamma_2}}
    ,\end{equation}
    where $L_0$, is the luminosity at which the break occurs and $\gamma_1$, $\gamma_2$ are the slopes of the power-law distributions below and above $L_0$.
    
\begin{figure*}
    \centering
    \includegraphics[width=\linewidth]{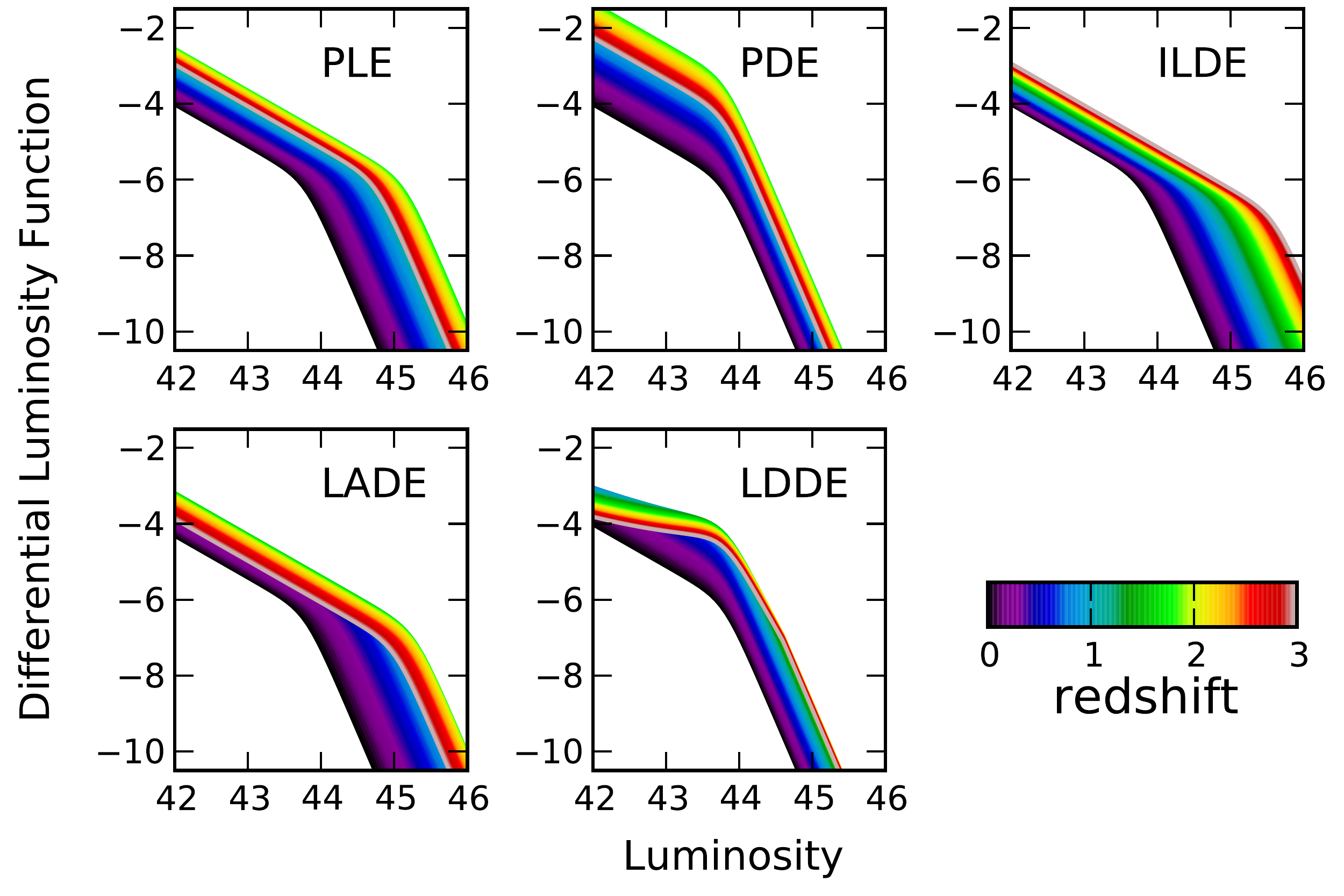}
    \caption[Luminosity function models]{Comparison of common AGN luminosity function evolutionary models. Models computed with fictitious evolutionary parameters to demonstrate the qualitative difference in evolution. The color scale shows the evolution of each model with redshift ranging from z=0 (black) to z=3 (red). The critical redshift is chosen $z_c=1.8$ (bright green) for all models.
    \label{fig:models}}
    \end{figure*}

Several works thereafter have concluded that the luminosity function shows a strong evolution with redshift \citep[e.g.,][]{Boyle1994,Page1996,Jones1997}. We test the most commonly used models to describe the evolution of the luminosity function. In the rest of this section we give the formula and a brief physical description of each model. In Fig. \ref{fig:models} we show a qualitative overview of the differential luminosity function versus the luminosity computed at several redshifts given by the color scale. In these plots the critical redshift after which the evolution changes has been set to $z_c=1.8$ when applicable.

        \subsection{Pure luminosity evolution}
        The first modification of the broken power-law model to include evolution, PLE, was examined by \citet{Schmidt1968}.
        The evolution is most commonly expressed as \citep{Miyaji2000, Ueda2003, LaFranca2005, Ebrero2009}

        \begin{equation}
        \frac{d\phi(L,z)}{d\log L} = \frac{d\phi(L/e(z),z=0)}{d\log L}
        \end{equation}
        with,
        \begin{equation}\label{eq:PLE}
        e(z) = 
            \begin{cases}
                (1+z)^{p_1} & z\le z_c\\
                (1+z_c)^{p_1}\cdot\left(\frac{1+z}{1+z_c}\right)^{p_2} & z\ge z_c\\
            \end{cases}
        ,\end{equation}
        where $z_c$ the redshift after which the evolution, $e(z)$, changes behavior and also follows a broken power law with slopes dependent on $p_1$ and $p_2$.  As seen from Fig. \ref{fig:models}, the PLE model is apparent as a shift of the luminosity function from higher to lower luminosities when moving from higher to lower redshift. Since the shape of the luminosity function is assumed to remain the same, this would be interpreted as a change in luminosity of the global AGN population. 

    \subsection{Pure density evolution} 
        The PDE model was also examined in the very early studies of luminosity function evolution \citep{Schmidt1968} and it is usually expressed as    
        \begin{equation}\label{eq:PDE}
        \frac{d\phi(L,z)}{d\log L} = \frac{d\phi(L,z=0)}{d\log L}\cdot e(z)
        ,\end{equation}
with the evolutionary factor, $e(z),$ given by Eq. \ref{eq:PLE}. The physical interpretation of this model is that AGN change in numbers, but their luminosities remain constant. This would be possible if the transition from active to inactive phase and vice versa were rapid and thus hardly observable. This evolution would appear as a change in the normalization of the luminosity function (see Fig. \ref{fig:models}). 
    
     \subsection{Independent luminosity density evolution} 
        The ILDE model was used by \citet{Yencho2009} to describe the evolution of AGN for redshifts below z=1.2. This model postulates that there is a simultaneous change in luminosity and number of AGN. Since this model is confined below redshift z<1.2, no critical redshift value was introduced, i.e.,\\

        \begin{equation}
            \frac{d\phi(L,z)}{d\log L} = \frac{d\phi(L/e_L(z),z=0)}{d\log L}e_D(z)
        \end{equation}
        with,
        \begin{equation}\label{eq:ILDE}
            e_L(z) = (1+z)^{p_L}
        \end{equation}
        and
        \begin{equation}
            e_d(z) = (1+z)^{p_D}
        ,\end{equation}
similar to the PLE (Eq. \ref{eq:PLE}) and PDE (Eq. \ref{eq:PDE}) models below $z_c$.

     \subsection{Luminosity and density evolution} 
        The LADE model was introduced by \citet{Aird2010}. This model enables independent evolution of the luminosity function both in luminosity and number density, but with the inclusion of a critical redshift value, 

        \begin{equation}\label{eq:fullLADE}
            \frac{d\phi(L,z)}{d\log L} = \frac{d\phi(L/e_L(z),z=0)}{d\log L}\cdot e_d(z)
        .\end{equation}
        The luminosity evolution follows a broken power law
        \begin{equation}\label{eq:LADE}
            e_L(z) = \left(\frac{1+z_c}{1+z}\right)^{p_1} + \left(\frac{1+z_c}{1+z}\right)^{p_2}
        ,\end{equation}
        while the evolution in number density follows a power law given by
        \begin{equation}
            e_d(z) = 10^{d(1+z)}
        .\end{equation}

    \subsection{Luminosity-dependent density evolution}  
       The LDDE model was introduced by \citet{Schmidt1983} to describe the evolution of optically selected quasars. \citet{Miyaji2000} introduced a formalism to describe the soft X-ray luminosity function of type 1 (unabsorbed) AGN, which has been extensively used ever since. This more complex model encapsulates the fact that the number density of AGN changes, but since the evolution of bright and low-luminosity AGN exhibits different timescales, the critical redshift, $z_c$, depends on the luminosity.

We use the formalism introduced by \citet{Ueda2003} as follows:
        \begin{equation}
        \frac{d\phi(L,z)}{d\log L} = \frac{d\phi(L,z=0)}{d\log L}\cdot e(L,z)
        \end{equation}
        with,
        \begin{equation}\label{eq:LDDE}
        e(L,z) = 
            \begin{cases}
                (1+z)^{p_1} & z\le z_c(L)\\
                (1+z_c)^{p_1}\cdot\left(\frac{1+z}{1+z_c}\right)^{p_2} & z\ge z_c(L)\\
            \end{cases}
        \end{equation}
        and 
        \begin{equation}\label{eq:zc}
        z_c(L) = 
            \begin{cases}
                z_c^{*} & L\ge L_a\\
                z_c^{*}\cdot\left(\frac{L}{L_a}\right)^a & L< L_a\\
            \end{cases}
        .\end{equation}
We express the evolution factor $e(L,z)$ of Eq. \ref{eq:LDDE} as
        \begin{equation}
        e(z, L) = \frac{(1+z_c)^{p_1}+(1+z_c)^{p_2}}{\left(\frac{1+z}{1+z_c}\right)^{-p_1}+\left(\frac{1+z}{1+z_c}\right)^{-p_2}}
        ,\end{equation}
        with $z_c$ defined as in Eq. \ref{eq:zc}. This formula is equivalent to the \citet{Ueda2003} evolution factor, creating a smooth transition of the XLF before and after the critical redshift, and it is normalized correctly at redshift zero.

\citet{Ueda2014}, motivated by the results obtained by the COSMOS team at high redshift, (XMM-COSMOS; \citet{Brusa2009}; Chandra-COSMOS; \citet{Civano2011}) introduced a more complicated model to describe the drop in number density above $\rm{z{\sim}3}$. In their formalism, an additional cutoff redshift is used, while the evolution index $\rm{p_1}$ is luminosity dependent as introduced by \citet{Miyaji2000}. In \citet{Ueda2014}, the evolutionary factor $\rm{e(L,z)}$ is given by
        \begin{equation}\label{eq:Ueda2014}
        e(L,z) = 
            \begin{cases}
                (1+z)^{p_1} & z\le z_{c1}(L)\\
                (1+z_{c1})^{p_1} \cdot \left(\frac{1+z}{1+z_{c1}}\right)^{p_2} & z_{c1}(L) < z\le z_{c2}(L)\\
                (1+z_{c1})^{p_1}\cdot \left(\frac{1+z_{c2}}{1+z_{c1}}\right)^{p_2}\cdot\left(\frac{1+z}{1+z_{c2}}\right)^{p_3} & z\ge z_{c2}(L)\\
            \end{cases}
        ,\end{equation}
        where 
        \begin{equation}\label{eq:Ueda2014zc}
        z_{c1,2}(L) = 
            \begin{cases}
                z_{c1,2}^{*} & L\ge L_{a1,2}\\
                z_{c1,2}^{*}\cdot\left(\frac{L}{L_{a1,2}}\right)^a_{1,2} & L< L_{a1,2}\\
            \end{cases}
        \end{equation}
        and
        \begin{equation}\label{eq:Ueda2014p1}
        p_1(L) = p_1^{*} + \beta_1 (\log{L}-\log{L_p})
        .\end{equation}
We  refer to this expression of LDDE hereafter as Ueda14.

\section{The $\rm{5-10\,keV}$ luminosity function}
\label{sec:results}
In this section, we present our procedure for the parameter estimation for the aforementioned models. We identify the model that best describes our dataset based on the Akaike information criterion (AIC) and Bayesian information criterion (BIC). 

\subsection{Model parameter estimation}
\label{sec:PamaterEstimation}

According to the Bayes theorem, the posterior probability of the model parameters $\theta$, given the observed data $D$ and the model $M$, $p(\theta|D,M)$, is proportional to the probability of prior knowledge of the model parameters $\theta$, $p(\theta|M)$, times the likelihood of observing the collected data $D$, under the given model $M$ and set of model parameters $\theta$, $p(D|\theta,M)$as follows:

\begin{equation}\label{eq:bayes}
    p(\theta|D,M) = \frac{ p(\theta|M)p(D|\theta,M) }{ \int p(\theta|M)p(D|\theta,M) d\theta }
.\end{equation}

Treating the observation of $n$ sources out of the available $N$ AGN in the Universe as a Poisson process\footnote{\citet{Kelly2008} provide an in-depth discussion on the Bayesian analysis of the luminosity function using a binomial distribution.} (applicable in the limit $n/N\ll1$) \citep{Marshall1983}, the likelihood\footnote{ Hereafter we use the symbol $\mathcal{L}$ to represent in the likelihood instead of $p(D|\theta,M)$ to be consistent with  XLF literature.} of observing $n$ sources is given by
\begin{equation}\label{eq:Likelihood}
\begin{split}
\ln\mathcal{L}(L,z)  = &\sum_{i=1}^n \ln\int\int{\phi(z_i,\log L_{xi})}p_i(\log{L}, z)\frac{dV_c}{dz}dzd\log{L} \\
& - \int\int\phi(z, \log L) \Omega(z, \log L)\frac{dV_c}{dz}dzd\log L
\end{split}
,\end{equation}
where $\phi(z,\log L)$ is the model of the luminosity function, $\rm{p(\log{L}, z)}$ the uncertainties on the luminosity and redshift for each observed source, $dV_c/dz$ the differential comoving volume, and $\Omega(z, \log L)$ the area curve of the survey.

Since X-ray counts and background counts are not available for all sources in our sample, we treat X-ray fluxes as delta functions. Hence, the uncertainties in Eq. \ref{eq:Likelihood} reduce to the uncertainties on the redshift estimation. In a similar fashion to \citet{Aird2010}, we introduce the full probability distribution function of the photometric redshift estimates in our computation. The photometric redshifts in our sample (XMM-COSMOS, LH, XMM-CDFS, and AEGIS-XD) were computed in a consistent way with special treatment for AGN, i.e., including hybrid templates and applying proper prior information on the absolute magnitude as defined in \citet{Salvato2011}.

We use noninformative distributions of priors for all parameters. The expected range for each parameter is determined from previous works on the AGN luminosity function in the $\rm{2-10\,keV}$ band, however, we do not wish to assume any shape for the distribution of the parameters.

To perform Bayesian analysis, we used MultiNest \citep{Feroz2008,Feroz2009,Feroz2013} through its python wrapper PyMultiNest\footnote{https://github.com/JohannesBuchner/PyMultiNest}\citep{Buchner2014}. MultiNest performs Nested Sampling introduced by \citet{Skilling2004} and  is able to explore the posterior even in the case of multimodal distributions.

\begin{figure*}
\begin{tabular}{cc}
\includegraphics[angle=0,width=0.45\linewidth, trim=10 30 10 40, clip]{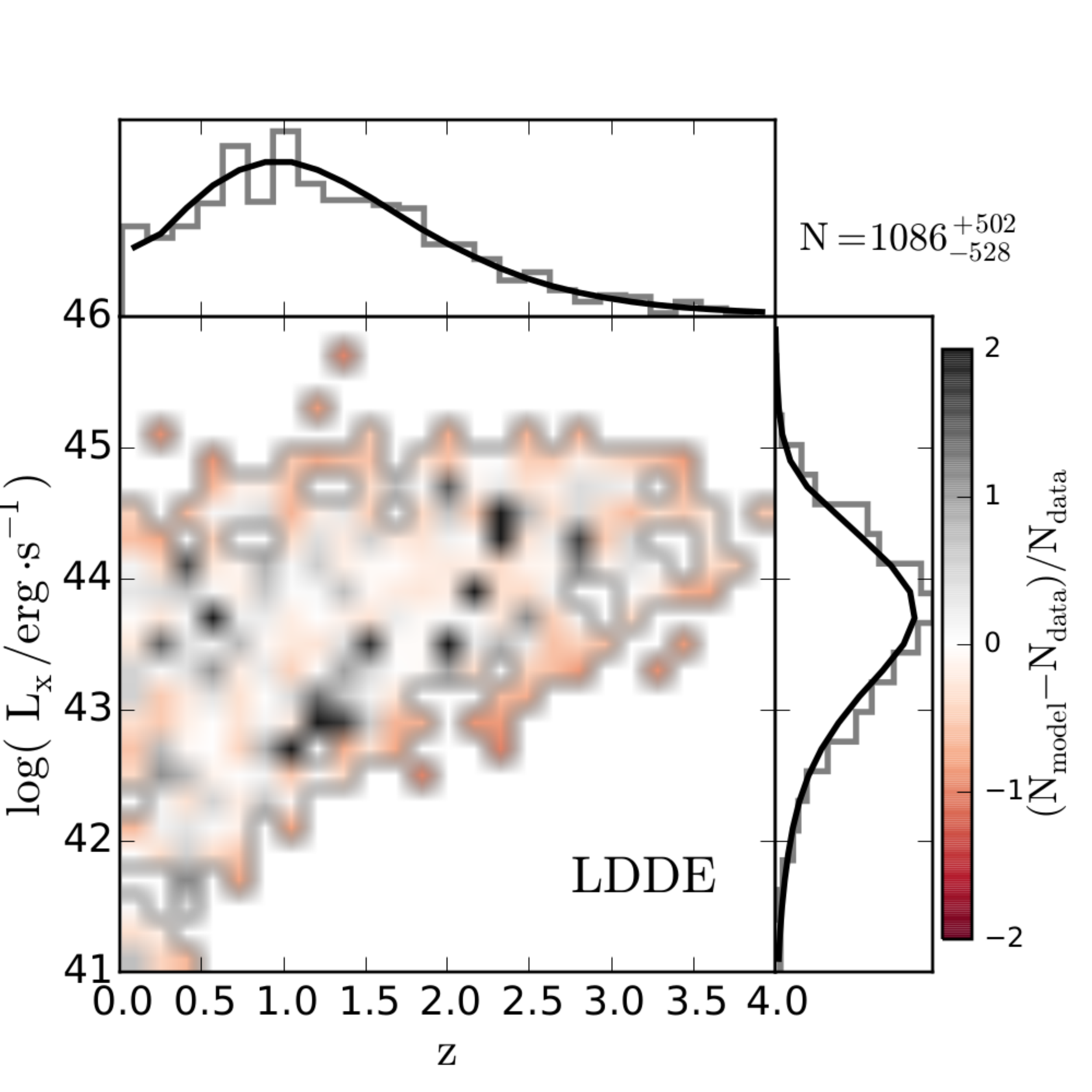}
&
\includegraphics[angle=0,width=0.45\linewidth, trim=10 30 10 40, clip]{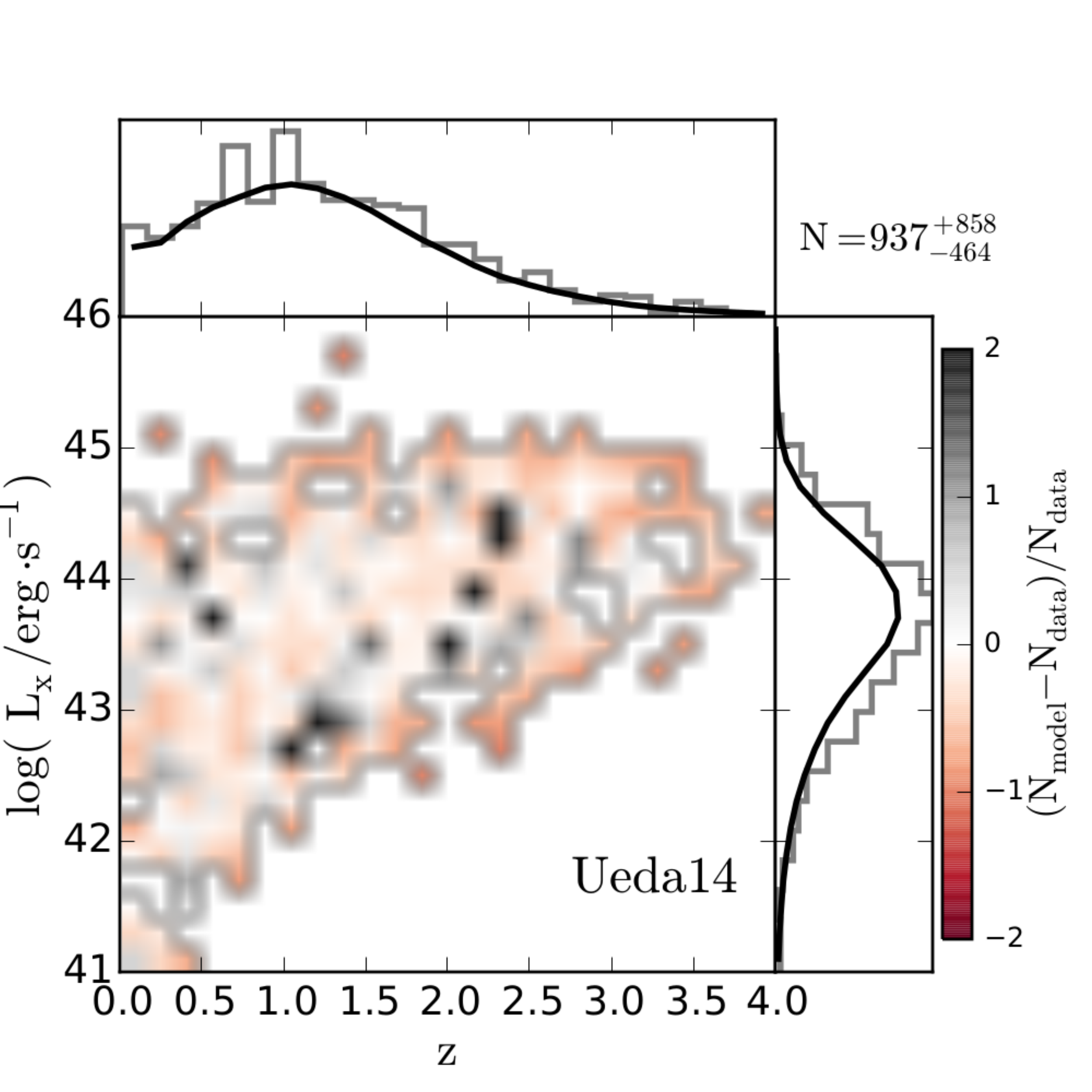}\\
\includegraphics[angle=0,width=0.45\linewidth, trim=10 30 10 40, clip]{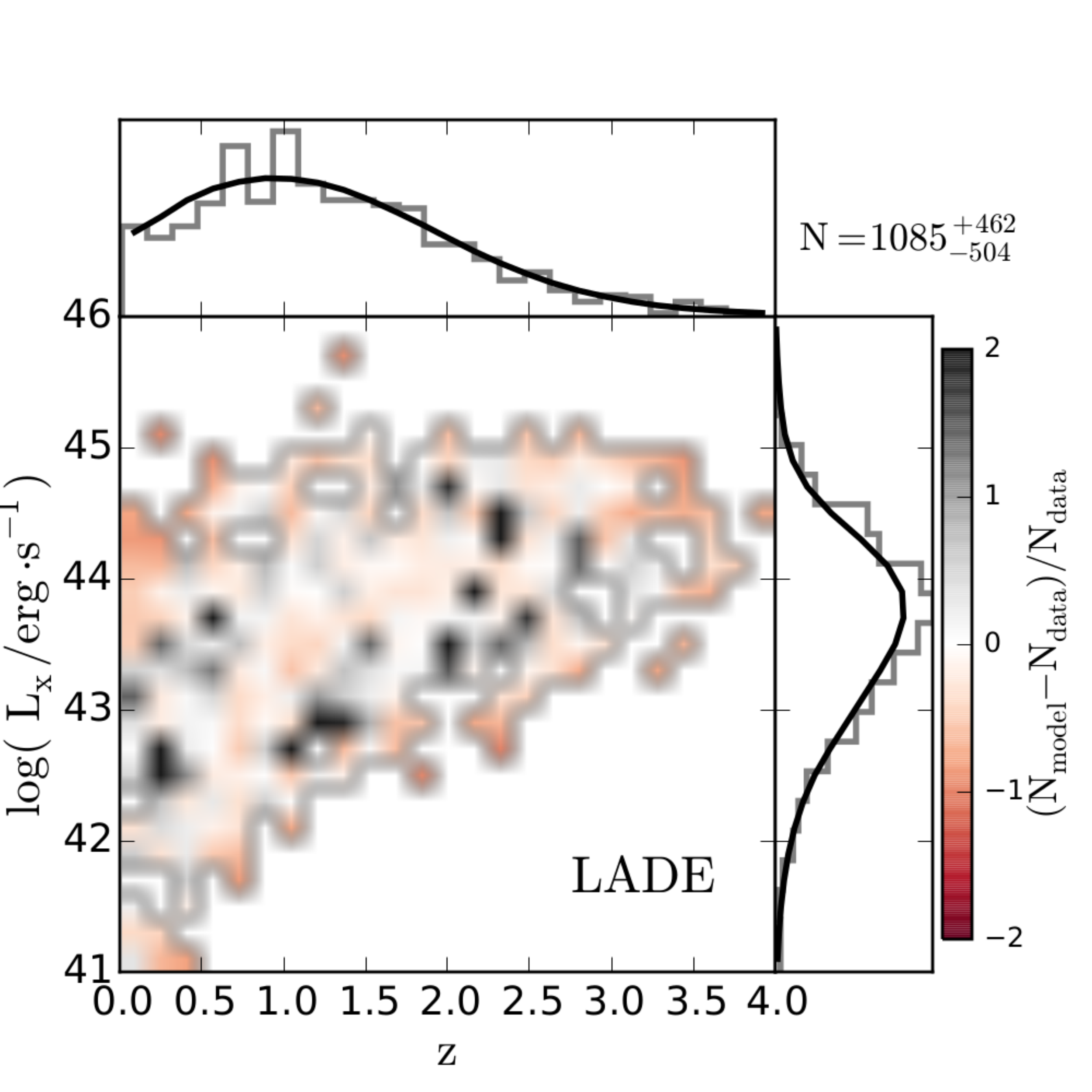}
&
\includegraphics[angle=0,width=0.45\linewidth, trim=10 30 10 40, clip]{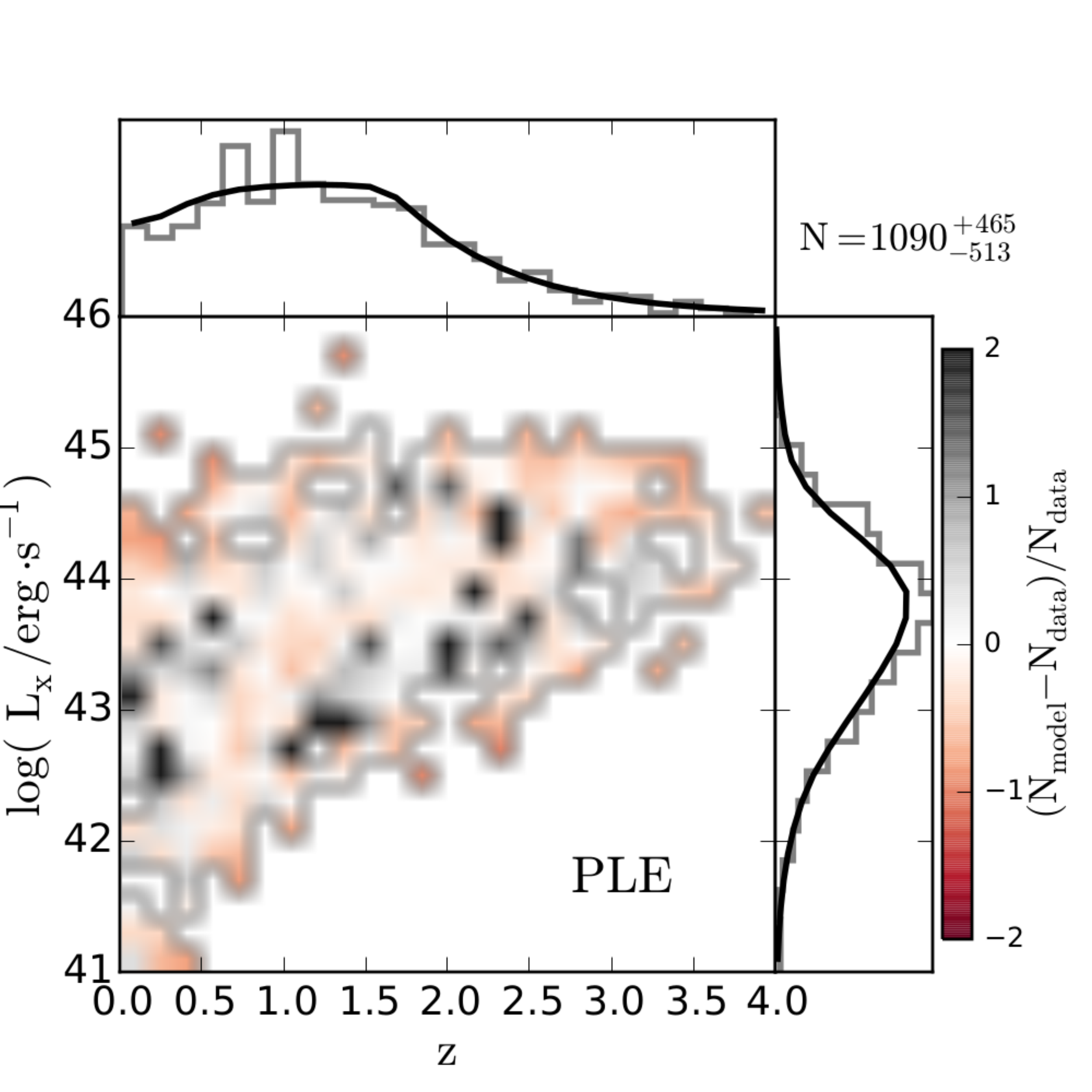}
\\
\includegraphics[angle=0,width=0.45\linewidth, trim=10 30 10 40, clip]{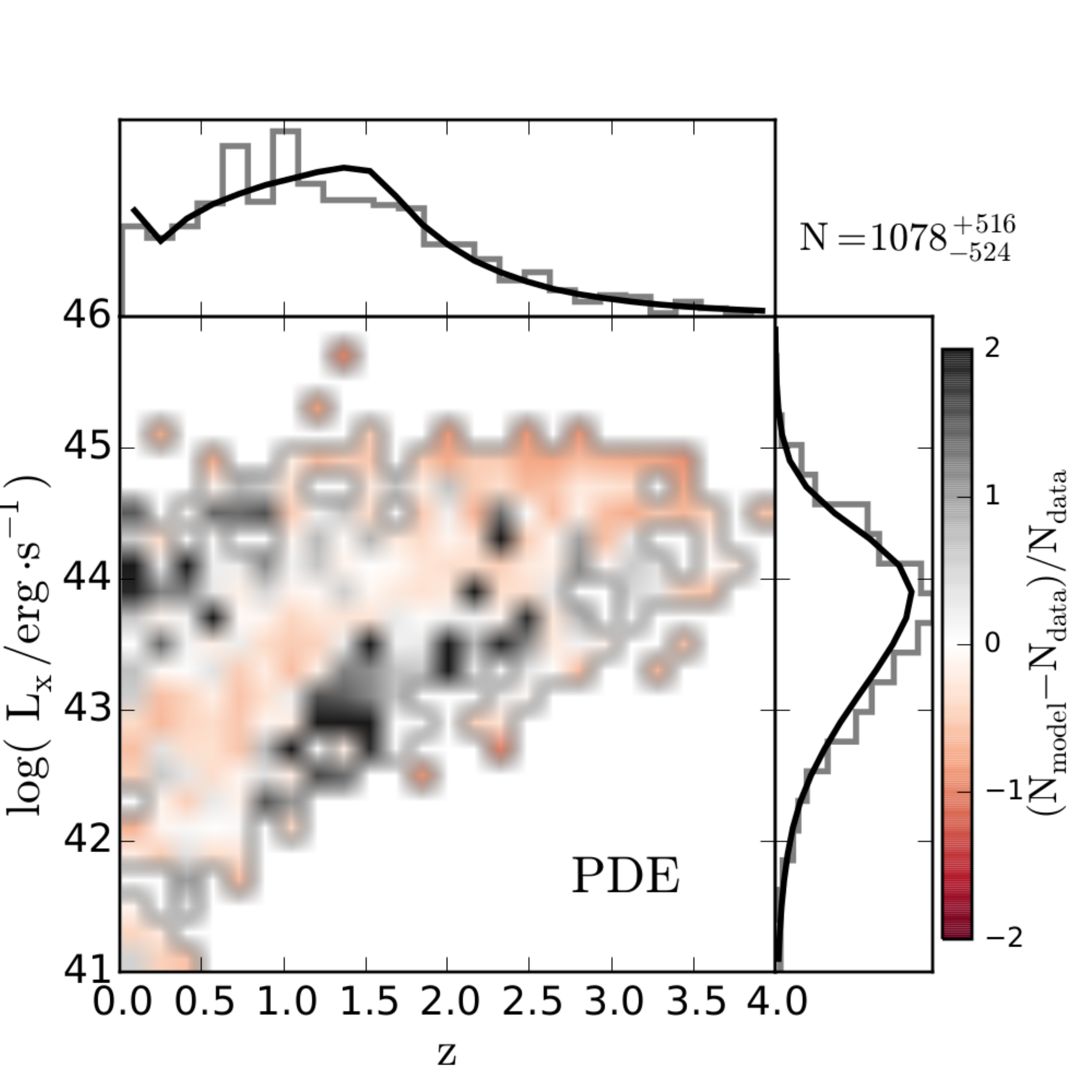} 
&
\includegraphics[angle=0,width=0.45\linewidth, trim=10 30 10 40, clip]{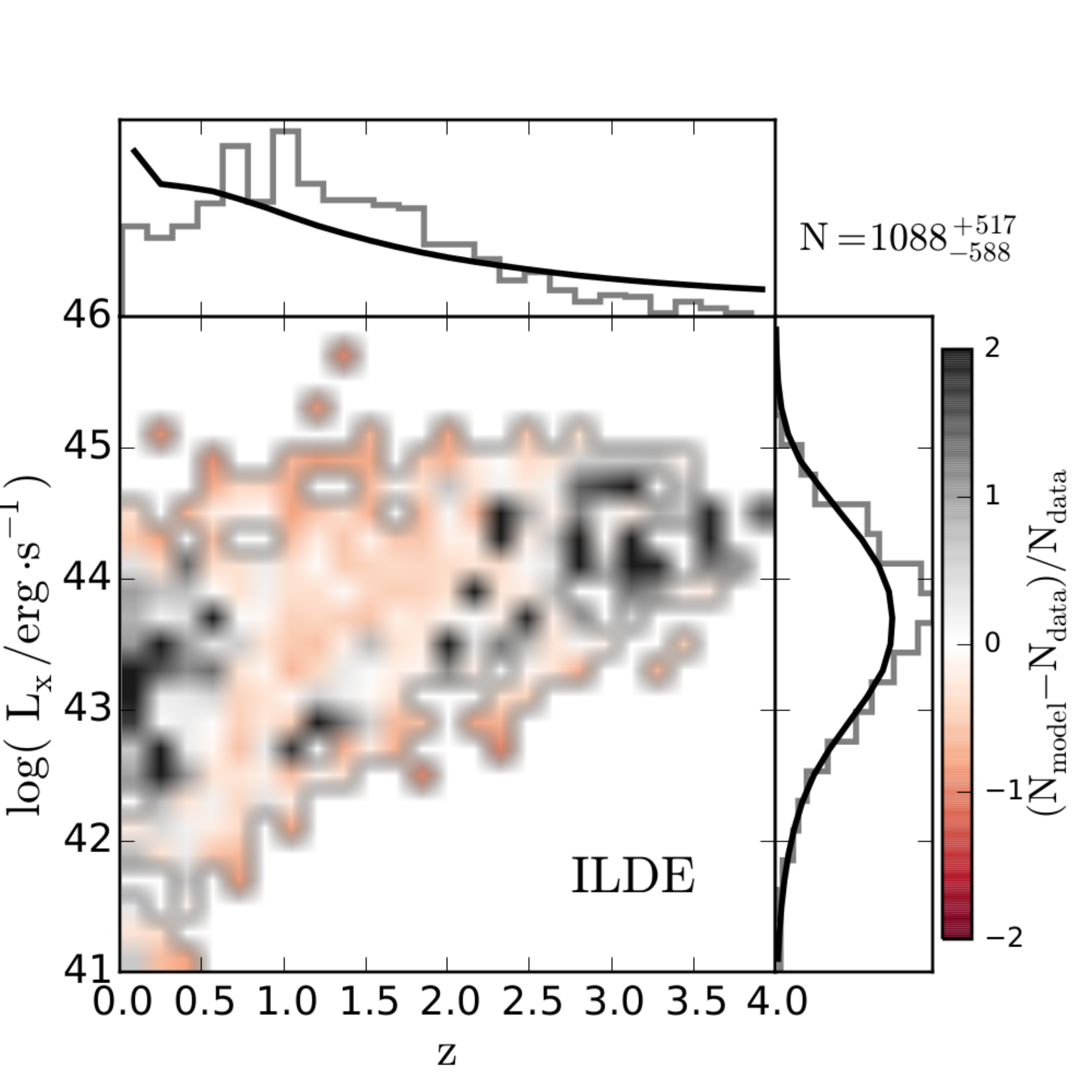}
\\
\end{tabular}
\caption{Fractional difference between predicted and observed number of AGN in a luminosity-redshift bin (color scale). The computation of the volume and luminosity function was performed on a 1000x1000 grid. The plots are rebinned on a 25x25 grid. Top and right-hand panels show the distribution in redshift and luminosity, respectively. The histograms represent the data, while the black lines are the model estimation.
\label{fig:posteriorcheck}}
\end{figure*}

 In Fig. \ref{fig:posteriorcheck} we plot the residuals between the model prediction and observed data for all models $(N_{model}-N_{data})/N_{data}$. All models are able to reproduce the total number of AGN observed (1105 objects available with redshift in our sample), albeit with about 40\%-50\% uncertainty at 90\% level. 
For each model, we also show the distribution in redshift (top panels) and in luminosity (right-hand panels). 
The gray histogram is the distribution of the observed data, while the solid black line is the prediction of each model. The redshift histograms indicate that the models do not capture the redshift spikes observed around $\rm{z=0.6}$ and $\rm{z=1.0}$. At low redshifts ($z<0.3$) we see that only the models LDDE and Ueda14 correctly capture  the local luminosity function. LDDE best reproduces   the one-dimensional distributions in redshift and luminosity and, at the same time,  the two-dimensional representation, while Ueda14 shows a suppression in the predicted number of sources. At low redshift, both LADE and PLE show an underestimated number of sources close to $\rm{\log L_x}=43$ and an overestimated number of sources at low ($\rm{\log L_x}=42$) and high ($\rm{\log L_x}=44$) luminosities.
Lastly, models PDE and ILDE appear unable to capture the change in number of AGN showing large islands of under- and overestimation of the number of AGN both at low and high redshifts.

\subsection{Model selection}
\label{sec:modelselection}

We apply the following two tests to identify the model that describes  our dataset best.

\paragraph{{\bf Akaike Information Criterion (AIC)}}

The AIC selects the model with the least information loss, penalizing models with a higher number of free parameters \citep{Akaike1974}. Therefore, it incorporates the Occam's razor, which among two equally plausible explanations prefers the simpler explanation. The AIC is computed as
\begin{equation}
{\rm{AIC}} = 2k -2\ln{\mathcal L}
\label{eq:AIC}
,\end{equation}
where $k$ is the number of parameters present in the model and $\mathcal L$ the maximum likelihood value. The preferred model is that with the lowest AIC value.

\paragraph{{\bf Bayesian Information Criterion (BIC)}}
The BIC selects models according to their likelihood value computed in a similar fashion to AIC, but with a higher penalty for complicated models \citep{Schwarz1978}. It is expressed as
\begin{equation}
{\rm{BIC}} = k\cdot\ln{n} -2\ln{\mathcal L}
\label{eq:BIC}
,\end{equation}
where $k$ is the number of parameters of the model and $n$ is the number of observations.
The preferred model is that with the lowest BIC value. Models with difference less than six must also be considered.

\begin{table}
\caption{Model selection criteria. According to AIC and BIC, LDDE is the preferred model. \label{tab:evidence}}
\begin{tabular}{l c c c c c c } \hline
                    & LDDE & Ueda14 & PLE & LADE & PDE & ILDE \\\hline
        $k$         &  9  &  15  & 7   & 8  & 7   & 6 \\ 
$\Delta$ AIC        &  0  &  13  & 41  & 37 & 150 & 269  \\
$\Delta$ BIC        &  0  &  43  & 31  & 32 & 140 & 254   \\ \hline
\end{tabular}
\end{table}

In Table \ref{tab:evidence} we give the comparison between LDDE and all other models. Both AIC and BIC identify the LDDE as the preferred model. For the dataset considered in this work, both selection criteria point to the direction that no other model can serve as an alternative. We see that according to the model selection criteria, ILDE is the worst representation; see  \S \ref{sec:PamaterEstimation} (Fig. \ref{fig:posteriorcheck}). From the simpler models, PDE performs worst than PLE. Additionally, we see that LADE, which allows an independent evolution in luminosity and number density, is comparable PLE.  Since BIC includes 15 model
parameters, it imposes a strong penalty on the LDDE formalism of \citet{Ueda2014} .

\begin{table}
\centering
\caption{LDDE parameter summary. We recommend the use of posterior draws sampled with MultiNest; available upon request. \label{tab:MultiNest}}
\begin{tabular}{c r@{.}lcr@{.}l r@{.}l r@{.}l}
\hline
\multirow{3}{*}{Parameter}  & \multicolumn{5}{c}{\multirow{3}{*}{Prior Interval}}   &           \multicolumn{4}{c}{Posterior Mode}  \\ \cline{6-10}
                            & \multicolumn{5}{c}{}                                  & \multicolumn{2}{c}{\multirow{2}{*}{Mean}}      & \multicolumn{2}{c}{Standard}  \\ 
                            & \multicolumn{5}{c}{}                                  & \multicolumn{2}{c}{}  & \multicolumn{2}{c}{Deviation}  \\ \hline
$\log L_0$  &  41&0  &--&  46&0  & 43&77 & \phantom{00}0&11  \\
$\gamma_1$  &  -2&0  &--&   5&0  &  0&87 & 0&06   \\ 
$\gamma_2$  &  -2&0  &--&   5&0  &  2&40 & 0&11   \\ 
$p_1$       &   0&0  &--&  10&0  &  5&89 & 0&31  \\ 
$p_2$       & -10&0  &--&   3&0  & -2&30 & 0&50  \\
$z_c$       &   0&01 &--&   4&0  &  2&12 & 0&16  \\ 
$\log L_a$  &  41&0  &--&  46&0  & 44&51 & 0&11  \\ 
$\alpha$    &   0&0  &--&   1&0  &  0&24 & 0&02 \\ 
$\log A $   & -10&0  &--&  -2&0  & -5&97 & 0&17  \\ \hline
\end{tabular}
\end{table}

\begin{figure*}
\center
\includegraphics[width=\textwidth]{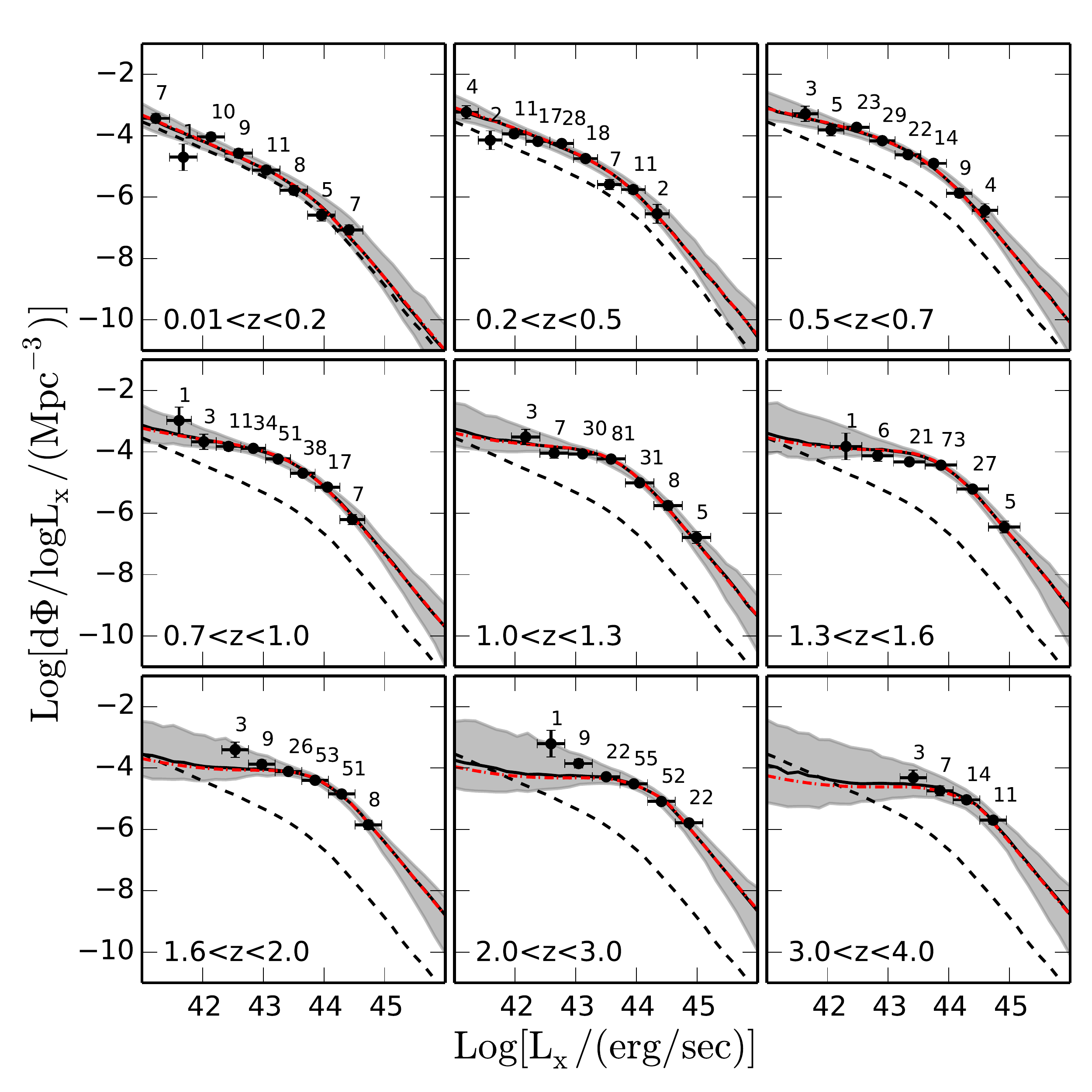}
\caption[LDDE Differential luminosity function]{Differential luminosity function versus luminosity for different redshift bins. The number of sources in each bin is given on the top right of each point. The solid black line is the mode of the luminosity function distribution, the gray shaded area encloses the 99\% credible interval. The dashed black line shows the estimated luminosity function at $\rm{z=0}$ for comparison. The two red lines show the estimated luminosity function when the two posterior modes are used separately. The points are the binned estimations of the luminosity function, according to the $\rm{1/V_{max}}$ method, and they are in excellent agreement with our model.
\label{fig:dPhi}}
\end{figure*}

\subsection{Results}

In Table \ref{tab:MultiNest} we summarize the parameter estimation of the LDDE model. We report  the uniform prior interval and the posterior mean and standard deviation  for each parameter. From Eq. \ref{eq:LF0} we can see that  parameters $\gamma_1$ and $\gamma_2$ are symmetric, which means that the posterior distribution contains two modes. These two modes are easily distinguished with MultiNest and summary statistics are provided for each mode (see also \S \ref{sec:perSurvey}).

In Fig. \ref{fig:dPhi} we plot the differential luminosity function for several redshift bins. The solid black line shows the peak of the distribution of the luminosity function at the median redshift of each redshift bin, while the gray shaded area encloses 90\% probability of the differential luminosity function, $d\phi/dlogL_x$. To determine this area, we compute $d\phi/dlogL_x$ in each redshift bin for 40 values of the luminosity, $L_x$, for all the draws from the posterior. In this way, we naturally incorporate  the true shape of the uncertainties for all parameters and their covariances. For reference, the dashed black line shows the luminosity function computed at redshift zero, using our model parameters in Eq. \ref{eq:LF0}. The two (overlapping) red lines show the luminosity function when mode 1 and mode 2 are considered separately. As expected and demonstrated in Fig. \ref{fig:dPhi}, the estimation of the LF is better constrained in the (L-z) locus where observations are available and less constrained when extrapolating\footnote{Assuming that the parameters are independent results in a large overestimation of the uncertainties on the luminosity function.}.

The black points are the result of the $1/V_{max}$ method. The $1/V_{max}$ method \citep{Schmidt1968} has been used widely in the literature to estimate the luminosity function of AGN. The advantage of this method is that the estimation is independent of any assumption on the underlying model. When a sizable sample of AGN is used, the computation of the luminosity function in thin redshift bins is powerful in revealing the presence of evolution. This computation also roughly reveals the shape of the XLF. Here we use the estimator proposed by \citet{Page2000}
    \begin{equation}\label{eq:PCVmax}
    \frac{d\phi(L,z)}{d\log{L}}=\frac{n}{\int_{\log{L_{min}}}^{\log{L_{max}}}\int_{z_{min}}^{z_{max}(L)}\Omega(z, \log L)\frac{dV}{dz}dz\,d\log{L}}
    ,\end{equation}
where $n$ is the number of AGN in the bin $[L_{min},L_{max}]$, and $[z_{min},z_{zmax}(L)]$. The value $z_{max}(L)$ corresponds to the maximum redshift up to which the $n$ sources would be still present in the sample, and this value is either the maximum redshift of the redshift bin or is given by the flux limit. Since it is a model independent method, it serves as a check for our fitting result.

\section{Discussion}
\label{sec:discuss}

\subsection{Information per survey}\label{sec:perSurvey}

\begin{figure*}
\centering
\includegraphics[width=\linewidth]{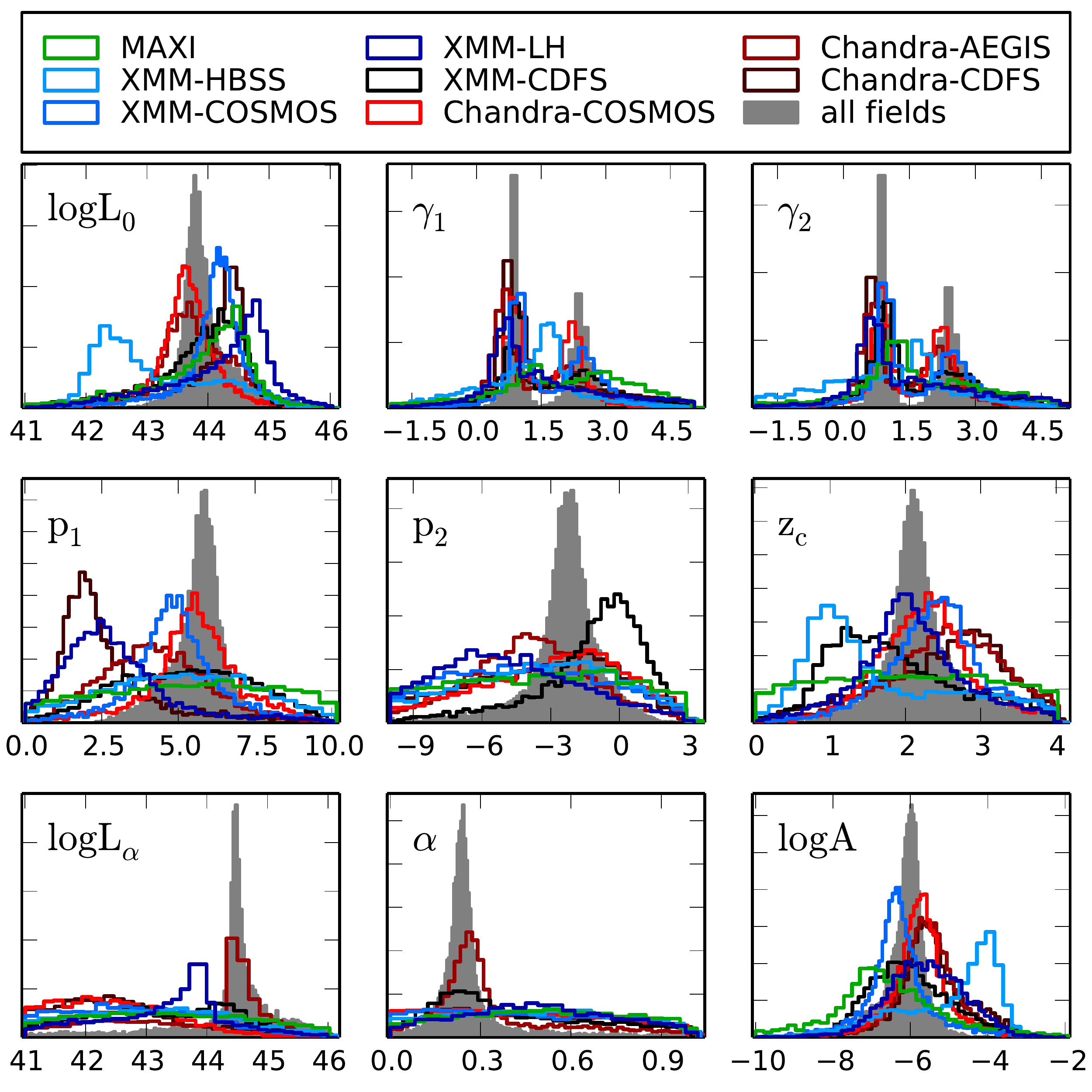}
\caption[Marginal distribution of LDDE model parameters]{Marginal probability distribution function for LDDE parameters per survey. The gray shaded area shows the distribution of the parameters when all the fields are considered. The colored lines show the luminosity function parameters determined separately in each field. Flat distribution means that no information is present in a given survey about a particular model parameter.
\label{fig:marginals}}
\end{figure*} 

In Fig. \ref{fig:marginals} we show the marginal posterior distribution of each parameter for the LDDE model. The gray shaded area represents the XLF parameter distribution when we take all of the surveys we used into account. The colored lines represent the LDDE parameters estimated from each survey independently. In a maximum likelihood estimation, a method commonly used in previous works, one would be forced to fix one or more evolutionary parameters  to gain some perspective on the local luminosity function when using a shallow survey such as MAXI.
One of the advantages of our analysis is exploring the posterior without having to assume any specific shape for the likelihood. All parameters are left free and any dataset can be used to extract information, even for complicated models. 
For example, as seen in Fig. \ref{fig:marginals} MAXI (green) and HBSS (light blue) carry essentially no information about $L_a$, and $a$. Therefore, the marginal distribution for these parameters has the same shape as the prior, which in our case was a flat distribution. On the other hand, the deepest fields AEGIS-XD (red), and XMM-CDFS (black) are able to put constraints on the evolution parameters.

We can quantify the information gained and encoded in the posterior distribution compared to the prior information using the Kullback-Leibler divergence \citep{KL1951}
\begin{equation}
{\rm{D_{KL}((P||Q)}} = \int_{-\infty}^{+\infty}p(x)\log_2\frac{p(x)}{q(x)}dx
\label{eq:DKL}
,\end{equation}
where P and Q are two continuous random variables and p, q are their corresponding probability density functions. The information gain is measured in bits since the logarithm with base 2 is used\footnote{One bit corresponds to the reduction of the standard deviation of a Gaussian distribution by a factor of three: http://www.mpe.mpg.de/\~jbuchner/bayes/utility/singlegauss\_bits.xhtml}. In Table \ref{tab:DKL} we show the information gain per survey and for combinations of surveys. In general, the shallow surveys provide higher information content on the LDDE parameters that describe the local XLF ($\rm{\log L_0}$, $\rm{\gamma_1}$, $\rm{\gamma_2}$) and the overall normalization ($\rm{A}$). On the other hand, deep surveys provide information on the evolution of the XLF with redshift ($\rm{p_1}$, $\rm{p_2}$). The luminosity dependence of the density evolution is constrained by the deepest fields, LH and AEGIS. The only field that does not conform to the expectation is Chandra-CDFS from which we find information gain for all model parameters, apart from parameter $\rm{\alpha}$.

Both astrophysical and systematic effects contribute to differences observed in the information present in each survey. Astrophysical effects include cosmic variance due to the presence of structures (e.g., clusters of galaxies), which would affect most the small-area fields (LH, XMM-CDFS, AEGIS). Also, because of the trade-off between depth and area covered by each survey, there is a lack of low-luminosity sources at higher redshifts. If a certain population  (type I vs II, or highly obscured AGN) is dominant at specific redshifts, this would bias our result. The biases introduced by astrophysical effects are effectively conquered by the combination of independent surveys.

Ideally, all data should be extracted in a homogeneous manner, but for technical reasons this is not always possible. Therefore, differences due to systematic effects include: 1) deviations in the calibration of the X-ray satellites; for instance, \citet{Lumb2001} and \citet{Tsujimoto2011} reported a 10-20\% difference in flux between XMM-Newton and Chandra; 2) difference in adopted survey detection limits determining the inclusion or not of a source in the catalog, which also affects the determination of the area curve of the survey; and 3) assumptions on the power-law index when converting counts to fluxes and fluxes to luminosities. We  tested the latter for the AEGIS-XD survey, comparing the evidence between $\Gamma=1.4$ and $\Gamma=1.9$. We find no difference in the preferred model in the two cases. The bottom panel of Table \ref{tab:DKL} shows the information gain when the surveys are considered in a coherent way \citep{Avni1980}. The combination of all XMM surveys provides enough information to constrain the behavior of the XLF at high redshifts through the model parameters $\rm{\log L_{\alpha}}$ and $\rm{\alpha}$. The Chandra surveys we consider cover a smaller area compared to the XMM surveys, but they are deeper that the XMM surveys. Runs 9 and 10 show the information gained when using only XMM surveys and only Chandra surveys, respectively. The parameter $\rm{p_1}$, which describes the evolution of the XLF below the critical redshift $\rm{z_c}$, is constrained better by the XMM surveys, while the high redshift evolution is constrained better by Chandra. In Table \ref{tab:XLFruns} we give the best-fit parameters for runs 9-14. There is notable difference between XMM-only and Chandra-only surveys. As seen from the information gain analysis, XMM surveys do constrain the high redshift evolution but with large uncertainties (20\% on parameter $\alpha$). On the other hand, Chandra-only fields lead to a somewhat biased estimation of the XLF evolution at low redshift (parameter $\rm{p_1}\approx7$). The reason for this discrepancy is the result of the area covered by the Chandra surveys, and not the result of systematic differences between the two instruments. At high redshift, both XMM and Chandra surveys converge to the same parameter estimation. The introduction of the MAXI survey significantly enhances the estimated XLF parameters, as seen by the information gained in Table \ref{tab:DKL}.

\begin{table*}
\centering
\caption{Prior-posterior information gain measured in bits on the marginal distribution of each model parameter. Zero bits means that the dataset carried no information regarding the specific parameter.\label{tab:DKL}}
\begin{tabular}{l c c c r@{.}l r@{.}l r@{.}l r@{.}l r@{.}l r@{.}l r@{.}l r@{.}l r@{.}l} 
\hline
\multirow{2}{*}{run} & \multicolumn{3}{l}{\multirow{2}{*}{field selection}} & \multicolumn{18}{c}{Information gain (bit)}\\ \cline{5-22}
 & \multicolumn{3}{c}{} & \multicolumn{2}{c}{$\log L_0$}  & \multicolumn{2}{c}{$\gamma_1$}  & \multicolumn{2}{c}{$\gamma_2$}  & \multicolumn{2}{c}{$p_1$}   & \multicolumn{2}{c}{$p_2$}  & \multicolumn{2}{c}{$z_c$} & \multicolumn{2}{c}{$\log L_a$}  & \multicolumn{2}{c}{$\alpha$} & \multicolumn{2}{c}{$\log A $}   \\ \hline 
1 & \multicolumn{3}{l}{MAXI}  &  0&67 & 0&26 &  0&40 & 0&04 &  0&03  & 0&02 & 0&05 & 0&04 & 0&53 \\
2 & \multicolumn{3}{l}{XMM-HBSS} &    0&54 & 0&66 &  0&38 & 0&13 &  0&10  & 0&30 & 0&13 & 0&04 & 0&76 \\
3 & \multicolumn{3}{l}{XMM-COSMOS} &   1&23 & 0&97 &  0&92 & 0&77 &  0&08  & 0&59 & 0&23 & 0&04 & 1&27 \\
4 & \multicolumn{3}{l}{XMM-LH} &    0&65 & 0&58 &  0&54 & 0&66 &  0&19  & 0&52 & 0&39 & 0&09 & 0&94 \\
5 & \multicolumn{3}{l}{XMM-CDFS} &    0&92 & 0&74 &  0&73 & 0&24 &  0&59  & 0&31 & 0&16 & 0&18 & 0&9 \\
6 & \multicolumn{3}{l}{Chandra-COSMOS} &   1&26 & 1&2 & 1&08 & 0&73 &  0&16  & 0&63 & 0&47 & 0&05 & 1&38 \\
7 & \multicolumn{3}{l}{Chandra-AEGIS-XD}  &  0&88 & 0&79 &  0&93 & 0&50 &  0&31  & 0&40 & 0&41 & 0&60 & 1&28 \\
8 & \multicolumn{3}{l}{Chandra-CDFS} &   0&83 & 0&72 &  0&68 & 0&91 &  0&13  & 0&28 & 0&35 & 0&05 & 1&13 \\ \hline
9 & -- & XMM & -- &   1&54 & 1&22 &  1&40 & 1&17 &  0&53  & 0&74 & 0&35 & 0&58 & 1&63 \\
10 & -- & -- & Chandra &   1&51 & 1&34 &  1&24 & 0&35 &  0&89  & 0&83 & 0&63 & 0&95 & 1&91 \\
11\tablefootmark{*} & -- & XMM & Chandra &    1&81 & 1&52 &  1&57 & 1&10 &  1&02  & 0&97 & 0&87 & 1&25 & 2&09 \\
12 & MAXI & XMM & -- &   1&65 & 1&33 &  1&29 & 1&59 &  0&58  & 0&74 & 0&38 & 0&71 & 1&67 \\
13 & MAXI & -- & Chandra &    1&63 & 1&39 &  1&31 & 1&31 &  1&09  & 0&74 & 0&78 & 1&27 & 1&82 \\
14\tablefootmark{*} & MAXI & XMM & Chandra  &    1&88 & 1&57 &  1&52 & 1&59 &  1&15 & 1&01 & 1&04 & 1&53 & 2&02 \\
\hline
\end{tabular}
\tablefoot{\tablefoottext{*}{In the runs where both XMM and Chandra fields are used, the overlap of the COSMOS and CDFS fields have been taken into account, as described in \S\ref{sec:data}.}
}
\end{table*}

\begin{table*}
\centering
\caption{LDDE parameter summary for runs 9-14.  \label{tab:XLFruns}}
\begin{tabular}{c r@{.}l r@{.}l r@{.}l r@{.}l r@{.}l r@{.}l }
\hline
                Parameter            &       \multicolumn{2}{c}{XMM}      &        \multicolumn{2}{c}{Chandra}  &   \multicolumn{2}{c}{XMM+Chandra} &     \multicolumn{2}{c}{MAXI+XMM}      &        \multicolumn{2}{c}{MAXI+Chandra}   & \multicolumn{2}{c}{MAXI+XMM+Chandra}   \\  
 \hline
$\log L_0$  & 43&97 $\pm$ 0.19 &  43&62 $\pm$ 0.14  & \phantom{0}43&72 $\pm$ 0.12 & \phantom{0}44&04 $\pm$ 0.15 & \phantom{00}43&77 $\pm$ 0.13 & \phantom{0000}43&77 $\pm$ 0.11\\
$\gamma_1$  &  0&97 $\pm$ 0.07 &   0&92 $\pm$ 0.09  &  0&86 $\pm$ 0.06 &  0&99 $\pm$ 0.07 &  0&86 $\pm$ 0.06 &  0&87 $\pm$ 0.06 \\ 
$\gamma_2$  &  2&53 $\pm$ 0.22 &   2&46 $\pm$ 0.17  &  2&37 $\pm$ 0.11 &  2&65 $\pm$ 0.22 &  2&43 $\pm$ 0.16 &  2&40 $\pm$ 0.11\\ 
$p_1$       &  5&72 $\pm$ 0.51 &   7&02 $\pm$ 0.93  &  6&03 $\pm$ 0.45 &  5&72 $\pm$ 0.36 &  5&92 $\pm$ 0.38 &  5&89 $\pm$ 0.31 \\ 
$p_2$       & -2&72 $\pm$ 1.14 &  -1&81 $\pm$ 0.56  & -2&19 $\pm$ 0.54 & -2&72 $\pm$ 1.07 & -2&19 $\pm$ 0.56 & -2&30 $\pm$ 0.50 \\
$z_c$       &  2&19 $\pm$ 0.35 &   1&99 $\pm$ 0.21  &  2&08 $\pm$ 0.17 &  2&28 $\pm$ 0.36 &  2&15 $\pm$ 0.25 &  2&12 $\pm$ 0.16 \\ 
$\log L_a$  & 44&55 $\pm$ 0.34 &  44&48 $\pm$ 0.10  & 44&49 $\pm$ 0.11 & 44&68 $\pm$ 0.34 & 44&53 $\pm$ 0.17 & 44&51 $\pm$ 0.11 \\ 
$\alpha$    &  0&26 $\pm$ 0.05 &   0&26 $\pm$ 0.03  &  0&25 $\pm$ 0.02 &  0&24 $\pm$ 0.04 &  0&24 $\pm$ 0.02 &  0&24 $\pm$ 0.02 \\ 
$\log A $   & -6&26 $\pm$ 0.28 &  -6&04 $\pm$ 0.24  & -5&92 $\pm$ 0.18 & -6&38 $\pm$ 0.24 & -5&98 $\pm$ 0.20 & -5&97 $\pm$ 0.17 \\ \hline
\end{tabular}
\end{table*}

\subsection{Luminosity function evolution}
Previous works in the $\rm{2-10\,keV}$ band have shown that the evolution of the AGN luminosity function is best described by the LDDE model \citep{Ueda2003, LaFranca2005, Silverman2008, Ebrero2009, Yencho2009, Ueda2014, Miyaji2015, Aird2015}.
The comparison among these works and the work presented here is not straightforward, as the sample selection and treatment vary. For example, each work treats  corrections for intrinsic absorption and corrections for redshift incompleteness differently. Additionally, not all models are tested in the literature for compliance with each dataset, and the most commonly used are the PLE and LDDE models. 

With this work we treat most of the analysis shortcomings found in earlier literature. We take  photometric redshift uncertainties for all sources into account, assuming a flat redshift probability distribution between $z=0-7$ for sources with no redshift estimation. As discussed in \S \ref{sec:perSurvey}, we do not fix any of the evolutionary parameters, since it is not needed within the Bayesian framework adopted for our analysis. Also, the $\rm{5-10\,keV}$ energy band chosen allows us to estimate the luminosity function of AGN, avoiding assumptions about the obscuration based on hardness ratios that could lead to a biased estimation of the absorbing column density \citep{Brightman2012}.

\begin{figure*}
\centering
\begin{tabular}{cc}
\includegraphics[width=0.49\linewidth]{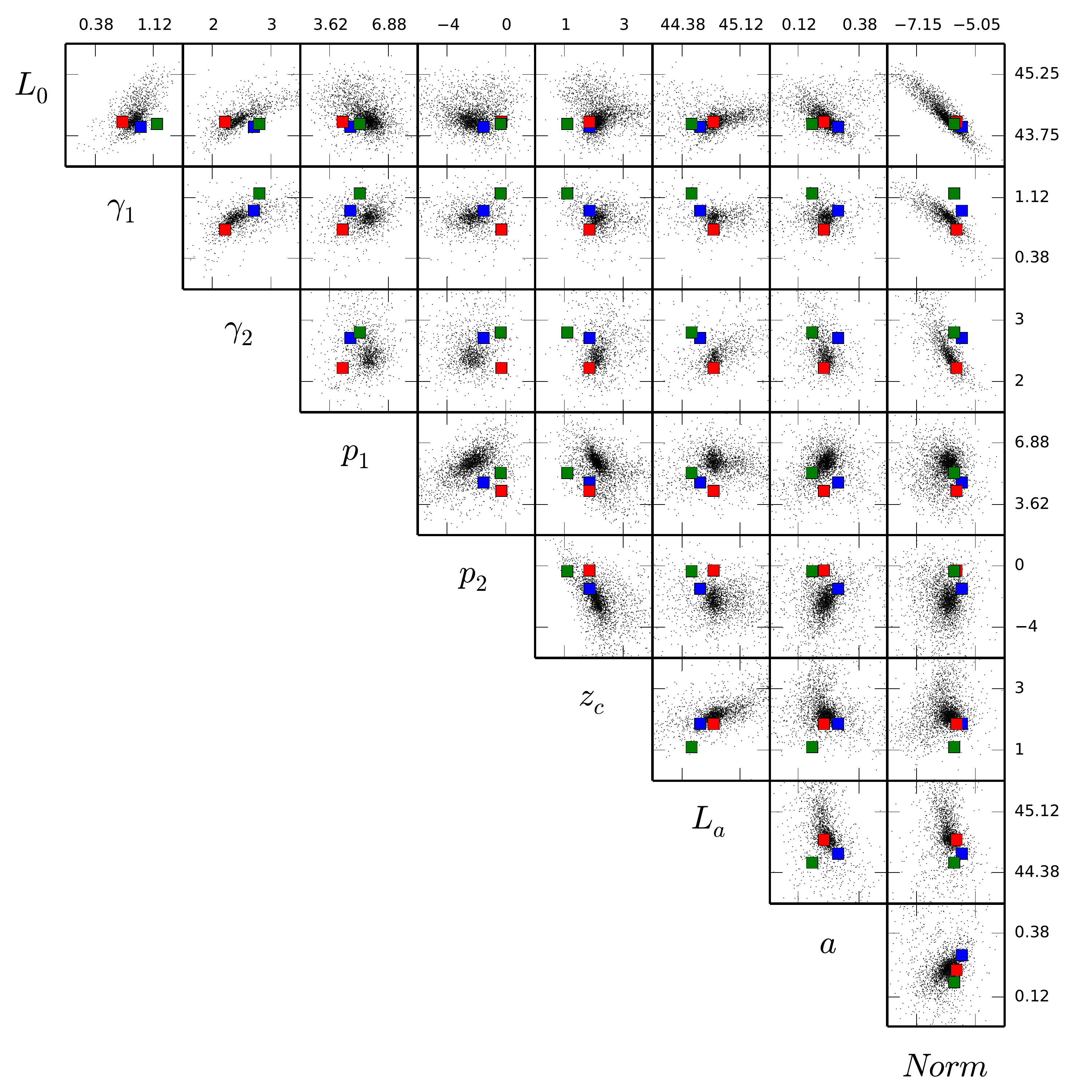} & \includegraphics[width=0.49\linewidth]{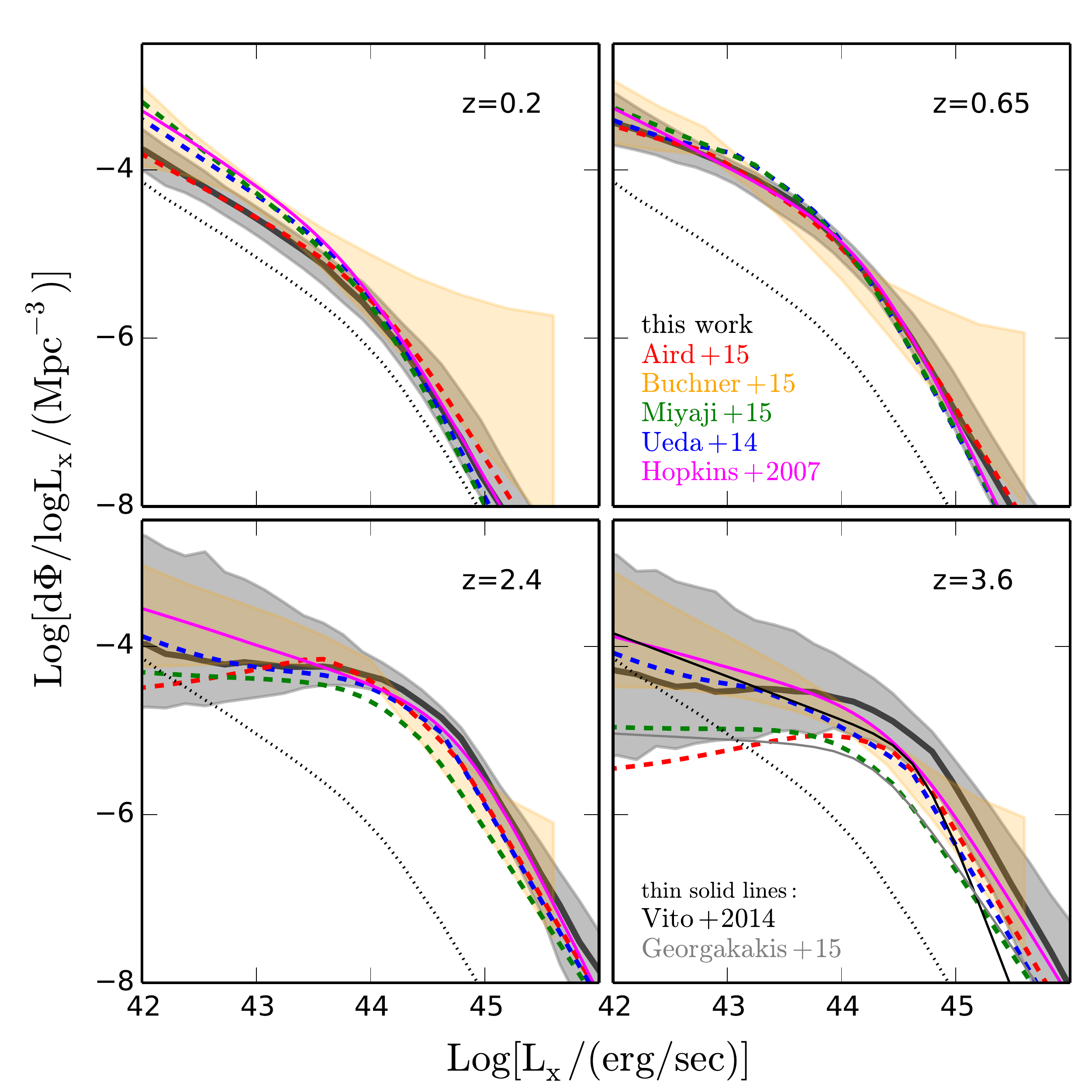}  \\
(a) & (b)
\end{tabular}
\caption{(a) Posterior draws from MultiNest (black) and comparison with the point estimates from \citet{Ueda2014} (blue), \citet{Miyaji2015} (green), \citet{Aird2015} (red). Comparison of $\rm{5-10\,keV}$ luminosity function (black line) transformed to $\rm{2-10\,keV}$ to literature results. The gray area shows the 90\% credible interval. Our result is in excellent agreement with the literature up to $z=2.5$. At high redshift our result shows a less prominent drop in the faint end of the XLF compared to estimations from \citet{Miyaji2015} (green), \citet{Aird2015} (red), and \citet{Georgakakis2015} (thin black line). \label{fig:compXLF}}
\end{figure*}

In Fig. \ref{fig:compXLF} (a) we provide a visualization of the MultiNest draws for our best model (black points). There are significant correlations between the model parameters that must not be neglected during the model parameter estimation, for example, by fixing model parameters. We also plot on the same figure the best parameter point estimates of \citet{Ueda2014} (blue), \citet{Miyaji2015} (green), and \citet{Aird2015} (red). We see that all model estimations agree on the break luminosity $\rm{\log L_0}$, the normalization and high redshift $\rm{z_c}$ evolution given by $\rm{\log L_{\alpha}}$ and $\alpha$ parameters. On the other hand, there is disagreement on the slopes of the XLF and their evolution with redshift. The surveys we use for all of the analyses contain strong evidence to constrain the break points in the XLF model, namely $\rm{\log L_0}$, $\rm{z_c}$, $\rm{\log L_{\alpha}}$. The exact shape, however, depends upon the quality of the redshift measurement, spec-z vs photo-z, and if the full PDF is taken into account. Additionally, the presence of absorption is an extra modeling factor that can introduce bias in the estimation of the XLF. By selecting objects in the $\rm{5-10\,keV}$ energy band, we remove the effects of this extra set of model assumptions.

In Fig. \ref{fig:compXLF} (b) we show our result transformed to the $\rm{2-10\,keV}$ energy band assuming AGN spectra can be described by a power law with $\Gamma=1.9$ (black solid line, dark gray area: 90\% credible interval) and  recent results from the literature (magenta, \citet{Hopkins2007}; blue, \citet{Ueda2014}; green, \citet{Miyaji2015}; yellow, \citet{Buchner2015}; red, \citet{Aird2015} ) for four redshift intervals. There is a remarkable agreement between all results up to redshift 2.5, and any small deviations are within the uncertainties. 

Despite the many efforts of deep X-ray programs, the AGN XLF above z=3 is still under debate. \citet{Brusa2009} noted that the high redshift number density of AGN in the XMM-COSMOS field showed a decline that is more rapid than the predictions of the XLF models. The same trend was found in the Chandra-COSMOS observations by \citet{Civano2011}. In the last panel of Fig. \ref{fig:compXLF} (z=3.6) we plot two more literature results from \citet{Vito2014} (thin black line) and \citet{Georgakakis2015} (thin gray line). Both works provide a parametric model estimation of the AGN XLF at 3<z<5. We see that the estimates cluster in two regimes: a significant drop in the number density \citep{Miyaji2015, Aird2015, Georgakakis2015} and a shallower drop in the number density \citep{Hopkins2007, Ueda2014, Vito2014, Buchner2015}. Our work is in agreement with the latter group and especially with the novel nonparametric XLF estimate of \citet{Buchner2015}. Even though \citet{Ueda2014} proposed a more complicated formalism for the LDDE model to take the change of the AGN number density at $\rm{z>3}$ into account, our work shows that at least up to redshift four, there are a set of model parameters able to describe the AGN XLF without introducing extra complexity in the LDDE model.

\subsection{AGN number density}

\begin{figure}
\centering
\includegraphics[width=0.95\linewidth]{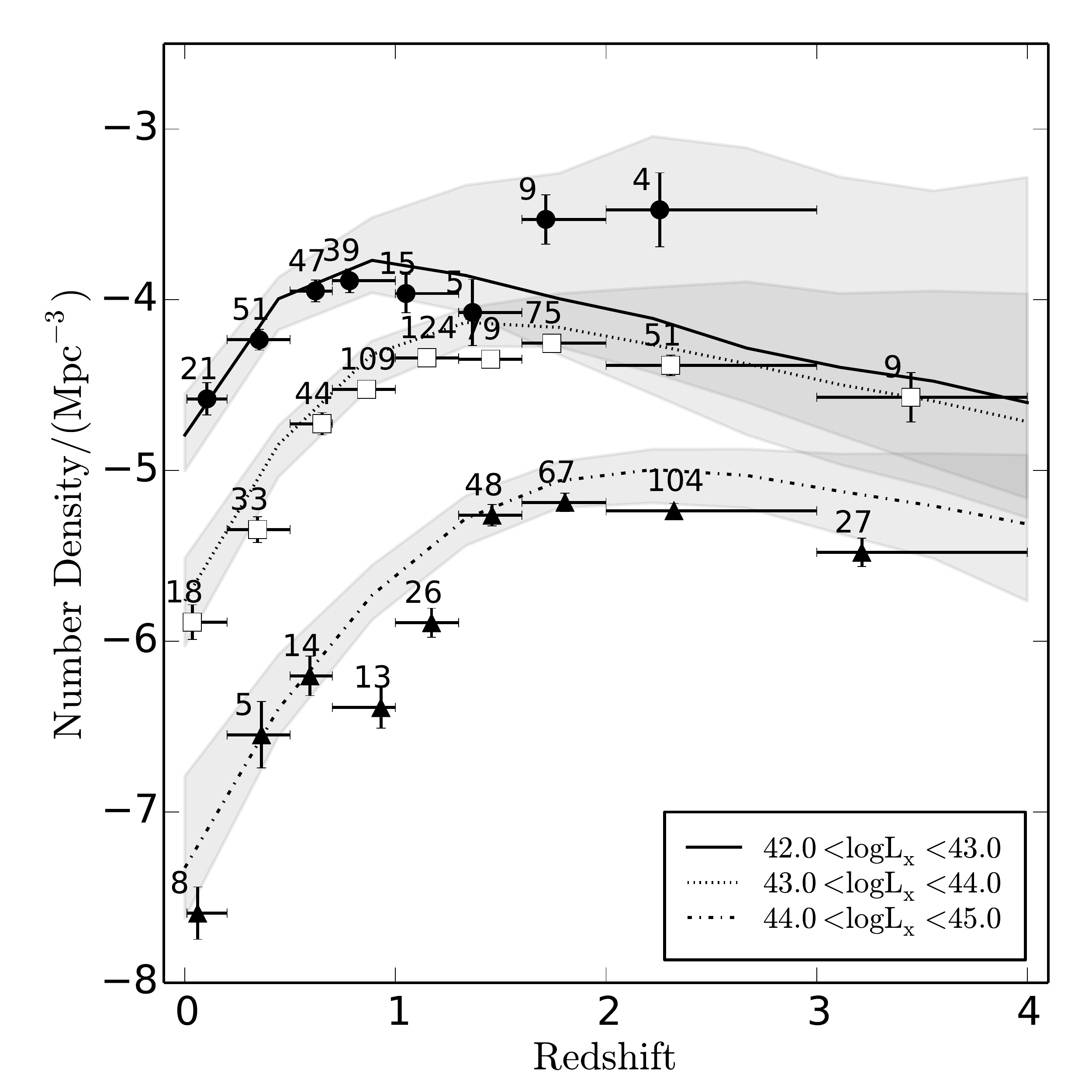}
\caption{Number density as a function of redshift for three luminosity bins showing the antihierarchical growth of black holes. The number of AGN per bin are given above each point.\label{fig:Ndensity}}
\end{figure}

In Fig. \ref{fig:Ndensity} we plot the number density as a function of redshift for three luminosity bins.
Our dataset shows the antihierarchical growth of black holes that has been observed previously in soft ($\rm{0.5-2\,keV}$) and hard ($\rm{2-10\,keV}$) energy bands. At the brightest luminosity range ($\rm{44<\log L_x<45}$), the number density for this dataset shows more of a flattening  than a drop, in contrast to the estimations of \citet{Brusa2009}, \citet{Civano2011}, and \citet{Kalfountzou2014}. \citet{Ueda2014} fix their second critical redshift at $\rm{z=3}$. \citet{Miyaji2015} use a sample of 3200 AGN in the $\rm{2-10\,keV}$ band and demonstrate that a second break is present between $\rm{z=2-3}$ after which a decline in the number density is observed. Our dataset is not numerous enough at $\rm{z>3}$ to support a more complicated evolution for AGN.  

The antihierarchical growth of supermassive black holes, as imprinted on the AGN luminosity function, might seemingly be in stark contrast with the hierarchical structure formation within the framework of $\Lambda$CDM cosmology. Through hydrodynamic simulations and semi-analytic modeling, it has been shown that it is possible to qualitatively reconcile structure formation and AGN activity. Nevertheless,  semi-analytic models are not able to rule out AGN evolutionary models or pinpoint the exact mechanism behind downsizing at this stage. Still largely debated in the literature, the predicted AGN luminosity function, from hydrodynamic simulation and semi-analytic modeling, is subject to the adopted AGN light curve and lifetime, obscuration, triggering mechanism, and feedback.

For example, in \citet{Fanidakis2011} the authors assumed that accretion takes place in two distinct modes, radiatively efficient and radiatively inefficient mode. Combining their semi-analytical model with the obscuration prescription from \citet{Hasinger2008}, they show that the bright end of the luminosity function is populated by quasars emitting close to the Eddington limit, while the faint end of the luminosity function is populated by black holes that undergo quiescent accretion.
\citet{Hirschmann2014} showed that their hydrodynamic simulation is able to reproduce the observed downsizing at redshifts $z=0-5  $, also allowing for cold gas accretion and major mergers. These authors mainly attribute the observed behavior to the gas density in the vicinity of the black hole, and they find the evolution of the gas reservoir was a consequence of star formation and AGN feedback. They also allow for obscuration in their models, following \citet{Hasinger2008}.

Even though dust obscuration can clearly enhance the observed flattening of the faint end of the luminosity function, which is particularly true for the soft X-ray band, the fact that downsizing is also observed in the $\rm{5-10\,keV}$ band, where the effect of the obscuration is minimal, points to the fact that this is not the predominant factor that shapes the AGN luminosity function. \citet{Hopkins2006} interpreted the AGN luminosity function in terms of quasar lifetimes and found good agreement between their simulations and the LDDE model, i.e., flattening of the faint end of the luminosity function. They claim that the observed break in the luminosity function corresponds to the maximum of the peak luminosity distributions of quasars at a certain redshift. The bright end of the luminosity function is populated by quasars that emit at their peak luminosity, while the faint end is populated by quasars that emit at lower luminosities. In their modeling, quasars spend the majority of their lifetime below their peak luminosity while, at the same time, more luminous objects transit to a less luminous stage faster than objects with lower peak luminosity. This implies that the slope of the faint end of the luminosity function is flatter at higher redshift. More recently, \citet{Enoki2014}, assuming only major merger driven AGN triggering and no obscuration effects, showed that a combination of parameters could reproduce the observed downsizing: 1) cold gas depletion due to star formation; 2) scaling of the AGN lifetime with the dynamical time, in which high redshift AGN has a shorter dynamical time; and 3) suppression of gas cooling in massive dark matter halos. 

\subsection{Future surveys}

With an anticipated detection of millions of AGN in the $\rm{0.5-2\,keV}$ energy range at a flux limit of $\rm{f_{0.5-2\,keV}=10^{-14}erg\,sec^{-1}cm^{-2}}$, eROSITA will mark a new era in the study of AGN. Over the $\rm{2-10\,keV}$ energy range, the brighter flux limit of $\rm{f_{2-10\,keV}=4\,10^{-14}erg\,sec^{-1} cm^{-2}}$ will allow the detection of hundreds of thousands of AGN \citep{Merloni2012}.

\begin{figure}
\centering
\includegraphics[width=\linewidth]{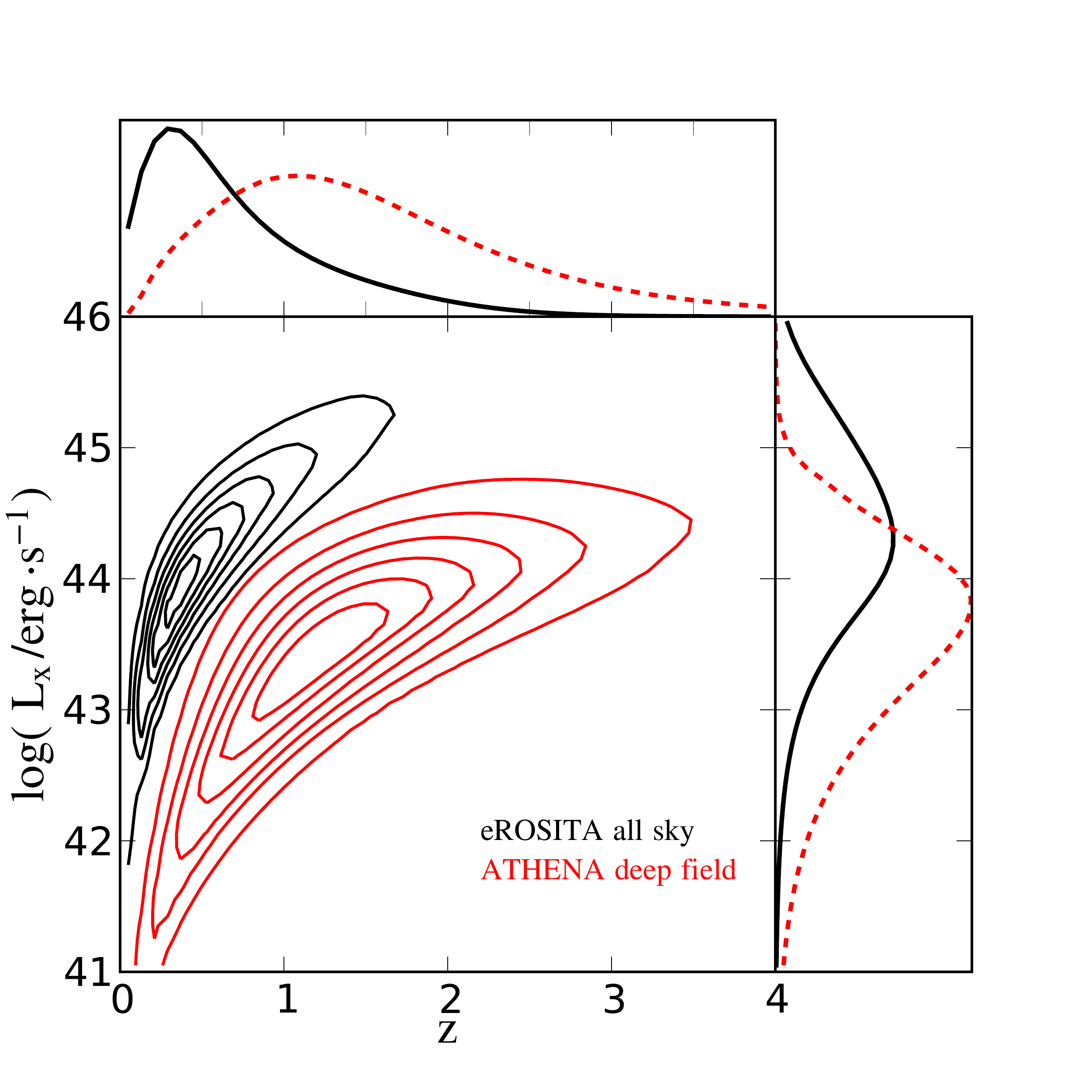}\caption{Expected luminosity-redshift plane coverage from the all-sky survey with eROSITA (black curves) and ATHENA (red curves) in the $\rm{2-10\,keV}$ energy band. Top panel shows the predicted redshift distribution. Panel on the right shows the predicted luminosity distribution. The two missions complement each other and both are pivotal for a precise estimation of the AGN luminosity function. \label{fig:eROSITA}}
\end{figure}

Transforming our $\rm{5-10\,keV}$ luminosity function in the $\rm{2-10\,keV}$ energy band and assuming an unabsorbed power law with photon index $\Gamma=1.9,$ we predict a total of $\rm{1.8 \cdot 10^5}$ unabsorbed AGN in the range $\rm{0.01<z<4}$ and $\rm{41<\log L_x<46,}$ assuming 80\% of the sky is accessible. This number is in good agreement with the prediction of \citet{Kolodzig2013}, which anticipated 130 000 AGN in the four-year survey over $\rm{34100\,deg^2}$. In Figure \ref{fig:eROSITA} (black contours) we show the expected number of AGN detected with eROSITA. The coverage of the L$_x$-z plane will not be homogenous, and the majority of the detected sources are close to the break of the luminosity function at redshift 0.5. The eROSITA all-sky survey will measure the local luminosity function of AGN and its evolution up to $\rm{z=1}$. At higher redshifts, constraints on the bright end of the luminosity function will be possible. The top and right-hand histograms show the expected number distributions of AGN in redshift and luminosity, respectively.

To determine the behavior of the AGN number density during the golden era of quasars at a comparable quality level to the sample provided by eROSITA, wide {\it and} deep surveys are needed. The ATHENA mission will be able to provide such observations. The red contours in Fig. \ref{fig:eROSITA} show the expected number of AGN from ATHENA, assuming the multitiered survey strategy described in \citet{Aird2013} and \citet{Georgakakis2013}, reaching a flux limit of $\rm{f_{2-10\,keV}=8\,10^{-17}erg\,sec^{-1} cm^{-2}}$ in the deepest area. The total anticipated number of AGN according to our model is $2.8\cdot10^5$ AGN in the redshift range $\rm{0<z<4}$ and $\rm{8\,10^4}$ AGN in the redshift range $1<z<2$, which is below the break luminosity $\log L_x<44$. With the combination of these unprecedented samples of AGN, we will be able to provide the most accurate constraints on the luminosity function and its evolution.

\section{Conclusions}
\label{sec:conclusions}
Combining the most recent X-ray observations from MAXI, HBSS, XMM-COSMOS, XMM-LH, XMM-CDFS, Chandra-COSMOS, Chnadra-AEGIS-XD, and Chandra-CDFS in the $\rm{5-10\,keV}$ energy band, we compile a sample of 1100 AGN with 98\% redshift completeness. We use the Chandra data in the inner region of the COSMOS and CDFS fields, and the XMM-Newton data in the outskirts where no Chandra data are available, to profit both from the depth and breadth of the observations. Our sample contains 68\% spectroscopic redshifts and 30\% very accurate photometric redshifts estimations from the fields XMM-COSMOS, XMM-LH, XMM-CDFS, Chandra-AEGIS-XD, Chandra-COSMOS, and Chandra-CDFS. Studying the $\rm{5-10\,keV}$ energy range, we avoid the potentially absorbed part of the spectrum for common $N_H$ values, effectively avoiding any assumption on otherwise necessary corrections to retrieve intrinsic X-ray luminosities.

Using Bayesian analysis we estimate the AGN $\rm{5-10\,keV}$ luminosity function and its evolution with redshift. Our results strongly support the luminosity-dependent density evolution (LDDE) model compared to PLE, PDE, LADE, and ILDE. We have demonstrated that by exploring the likelihood via MultiNest, we no longer need to fix model parameters even when there is not enough information in a dataset to constrain a complicated model. We quantify the information gain per XLF model parameter for the LDDE model for each individual field and for XMM and/or Chandra field combinations. We show that the evolution predicted by XMM-only and Chandra-only fields is varying, but this evolution remains consistent within the uncertainties at the 90\% level.

We rule out the possibility of a pure density evolution. Our results demonstrate the presence of one critical redshift after which the evolution changes behavior and shows no evidence of a second critical redshift up to redshift four. Future work should focus on physically motivated evolutionary models, coupling the observed change in number density with AGN physics. We rule out the possibility of obscuration as the primary explanation for the observed antihierarchical growth of AGN.

Planned X-ray observatories will give an unprecedented view of the bright end of the luminosity function. We predict $1.8\cdot10^5$ AGN up to $\rm{z=4}$ in the four-year, all-sky survey with eROSITA, while the multitiered survey strategy with ATHENA proposed in \citet{Aird2013} would provide the outstanding number of $10^4$ AGN at $1<z<2$ and $\log L_x<44$, which is sufficient to study the {\it golden era} of quasars and their coevolution with galaxies.

\begin{acknowledgements}
SF acknowledges a grant from the Swiss National Science Foundation. This work benefited from the THALES project 383549, which is jointly funded by the European Union and the Greek Government in the framework of the program "Education and lifelong learning". Financial contribution from the agreement ASI-INAF I/009/10/0 and “PRIN–INAF 2011" is acknowledged. MB acknowledges support from the FP7 Career Integration Grant ``eEASy'' (``SMBH evolution through cosmic time: from current surveys to eROSITA-Euclid AGN Synergies", CIG 321913). TM is supported by UNAM DGAPA Grant PAPIIT IN104113 and CONACyT Grant 179662.
\end{acknowledgements}

\bibliographystyle{aa} 

\end{document}